\documentstyle[12pt,a4]{article}
\newcommand{\vs}[1]{\rule[- #1 mm]{0mm}{#1 mm}}

\begin{document}
\renewcommand{\thefootnote}{\fnsymbol{footnote}}
\newpage
\pagestyle{empty}
\setcounter{page}{0}


\null
\begin{minipage}{4.9cm}
\end{minipage}
%

\begin{center}
{\Large {\bf THE PRODUCTION OF RESIDUAL NUCLEI IN}}\\[0,3cm]
{\Large {\bf PERIPHERAL HIGH ENERGY}}\\[0,3cm]
{\Large {\bf NUCLEUS-NUCLEUS INTERACTIONS}}

\vs{5}

{\bf A. Ferrari, P.R. Sala}\\
{\em INFN, Sezione di Milano, Via Celoria 16, I-20133 Milano, Italy}

\vs{0,5}

{\bf J. Ranft}\\
{\em Departamento de Fisica de Particulas, Universidade de Santiago 
     de Compostela,\\
     E-15706 Santiago de Compostela, Spain}

\vs{0,5}

{\em and}

\vs{0,5}

{\bf S. Roesler}\\
{\em Universit\"at Siegen, Fachbereich Physik, D-57068 Siegen, Germany}

\end{center}
\vs{5}

\centerline{ {\bf Abstract}}
\indent

A formation zone intranuclear cascade model is applied to peripheral
nucleus-nucleus collisions. We calculate the excitation energies of
prefragments, treat their further nuclear disintegration and
introduce a model for nuclear deexcitation by photon emission. Results
are compared to data on target associated particle production in
nucleus-nucleus collisions. We discuss implications of these models to
the description of particle production in the fragmentation regions.
Special emphasis is put on applications for air showers induced by
cosmic ray nuclei and for residual nucleus production at heavy ion
colliders.

\vfill
\rightline{Siegen SI 96-02}
\rightline{Santiago de Compostela US-FT/9-96}
\rightline{March 1996}

\newpage
\pagestyle{plain}
\renewcommand{\thefootnote}{\arabic{footnote}}
%
%
\section {Introduction}
\noindent
High energy nucleus-nucleus collisions can be described by multiple
scattering processes between nucleons of the interacting nuclei in which
most of the final state particles are produced. These particles either
escape directly from the interaction region or may induce intranuclear
cascade processes, i.e. they may ``knock out'' further nucleons while
penetrating through the spectator parts of the nuclei. Both, the primary
and secondary nucleon-nucleon interactions and the intranuclear cascade
contribute to the excitation of the spectator fragments. After peripheral
collisions their deexcitation proceeds via evaporating nucleons and light
fragments and via photon-emission until a stable configuration is
reached, which we call ``residual nucleus''. 
In contrast, for low impact parameters or small, highly excited
prefragments other nuclear disintegration processes, such as
multifragmentation, become important.

Primary and secondary interactions between nucleons of both nuclei
dominate the hadron production in most of the rapidity region covered by
the interaction and are well described in the framework of the Dual
Parton Model (DPM)~\cite{Capella94a}. They were extensively studied with
the Monte Carlo (MC) implementations of this model {\sc
dtunuc}~\cite{Moehring91} and {\sc dpmjet-ii}~\cite{Ranft94c}.
However, when dealing with particle or fragment production in the 
forward or backward fragmentation regions a detailed description of 
intranuclear cascade processes and of nuclear disintegration is of 
particular
importance. Examples are air showers induced by high energy cosmic ray
nuclei or aspects of cascade processes in matter initiated by heavy ions
in general. A formation zone intranuclear cascade model, the calculation
of nuclear excitation energies, and models for nuclear evaporation, 
high energy fission and the break-up of light nuclei were discussed
in~\cite{Ferrari95a}. It was shown that their MC-realizations 
for hadron-nucleus interactions describe successfully the basic features 
of target associated particle production.

In this paper, these models are applied to nucleus-nucleus 
collisions and extended by a model for $\gamma$-deexcitation. 
We stress, that the formation zone intranuclear cascade model and the
calculation of nuclear excitation energies as they are formulated and
presented here can only be expected to work reliably in
peripheral collisions. Nevertheless, we apply them in minimum bias
situations to all collisions, since the fraction of central collisions, 
where they might fail is rather small. It should definitely not be applied 
to central collisions.  Most of the data, we have to
compare with are for asymmetric collisions of typically one light
projectile nucleus with a heavier target nucleus. It seems, that the
model performs rather well in such collisions. We do however not know,
and there are so far no data to compare with, how well the model
performs in collisions of identical or nearly identical heavy nuclei.

In Sect.2 we summarize briefly the main steps of sampling nucleus-nucleus 
interactions within the event generators {\sc dtunuc 2.0} and 
{\sc dpmjet-ii}. Furthermore,
we summarize the basic ideas of the formation zone intranuclear cascade
model, describe the application to nucleus-nucleus collisions and
calculate nuclear excitation energies. In
Sect.3 a model for nuclear deexcitation by photon emission is introduced
and its MC-implementation is discussed. In Sect.4 the production of slow,
target associated particles is compared to experimental data.
Cross sections for the production of residual nuclei are calculated and
compared to measured cross sections. In Sect.5 we discuss applications to
air showers induced by cosmic ray nuclei and give cross sections for
residual nuclei production in interactions with air. Furthermore, mass
distributions of residual nuclei produced in lead-lead collisions at
RHIC-energies are discussed. Finally, in Sect.6 we summarize our results.
%
%
\section{The formation zone intranuclear cascade model for
         nucleus-nucleus collisions}
\subsection{The event generators {\footnotesize DPMJET-II} and 
            {\footnotesize DTUNUC 2.0}}
Both MC event generators, {\sc dpmjet-ii} and {\sc dtunuc 2.0}, 
start with sampling the spatial initial configuration, i.e.
the positions of the nucleons in space-time in the rest system of the
corresponding nucleus, from standard density distributions. The nucleons
are assumed to behave like a Fermi-gas. Therefore, the momenta assigned
to the nucleons are sampled from zero-temperature Fermi distributions. 

The collision proceeds via $\nu$ nucleon-nucleon interactions between
$\nu_p$ and $\nu_t$ nucleons from the projectile and target, resp. The
values $\nu, \nu_p, \nu_t$, and the impact parameter of the
nucleus-nucleus collision are obtained according to the Glauber-theory
using the MC-algorithm of~\cite{Shmakov89}. Particle production in each
nucleon-nucleon interaction is described by the DPM but based on
different MC-realizations in the two event generators. The {\sc dpmjet-ii} 
code includes the {\sc dtujet-93}~\cite{Aurenche94a}
event generator for hadron-hadron interactions, whereas in the
{\sc dtunuc 2.0} code a one-Pomeron exchange model for low energies or the 
{\sc phojet}~\cite{Engel95a,Engel95d,Engel95b} description of the DPM can 
be chosen optionally. The application of {\sc dpmjet-ii} and
{\sc dtunuc 2.0} is therefore restricted to upper energy
limits of about 43~TeV and 1~TeV\footnote{An extension to higher
energies is planned for the near future~\cite{Engel96pc}}$^)$~\cite{Engel95d}
c.m. energy per nucleon up to which the models {\sc dtujet} and 
{\sc phojet} are valid. For details we refer to the literature. 

Using a quasi-classical description, the hadrons
produced in the $\nu$ nucleon-nucleon interactions are not only known with
their 4-momentum but also with their position in space-time. They may
cause intranuclear cascade processes in the spectator parts of the
nuclei which were not involved in the primary and secondary
interactions. Both mechanisms of particle production, the $\nu$
nucleon-nucleon interactions and the intranuclear cascade contribute to
the excitation of the spectators. The parts of the code concerning the
formation zone intranuclear cascade, the calculation of excitation
energies and the various models treating the disintegration and
deexcitation of prefragments, which are subject to this work, are 
identical in {\sc dpmjet-ii} and {\sc dtunuc 2.0}. 
\subsection{\label{fzcAA} The formation zone cascade in the 
            spectator prefragments}
A description of how to implement the formation zone intranuclear
cascade~\cite{Ranft88a,Ranft89a} in hadron-nucleus collisions was given 
in~\cite{Ferrari95a}.
Therefore, in this Section we summarize only the basic ideas.

Within the quark model, the states being
formed in the primary nucleon-nucleon interaction can be understood as
consisting of valence quarks only, i.e without the full system of sea
quarks, antiquarks, and gluons and have therefore a reduced probability
for hadronic reinteractions inside the nucleus~\cite{Ranft88a}.
These reduced interaction probabilities can
be taken into account by assigning formation times to the hadrons
produced\footnote{In general one would have to consider cascade
processes initiated by resonances instead of those initiated by stable
hadrons. But since interaction cross sections of resonances are less
well known we start from hadrons and compensate the effects of this
approximation by using an effective formation time}. The hadrons are 
not able to reinteract with nucleons of the residual spectator nuclei 
within
that time. For each secondary a formation time $\tau$ in its rest system
is sampled from an exponential distribution with an average value $\tau_s$
defined by~\cite{Ferrari95a,Ranft89a}
\begin{equation}
\label{deffortim}
\tau_s=\tau_0\frac{m_s^2}{m_s^2+p_{s\perp}^2}.
\end{equation}
$m_s$ and $p_{s\perp}$ are the mass and the transverse momentum of the
secondary and $\tau_0$ is a free parameter which has to be adjusted by
comparing particle production within the model to experimental data.
We fix $\tau_0$ at a value of 1.9~fm/$c$.
After having assigned a formation time to a secondary its spatial
positions in the rest system of both nuclei are known and we start
with considering an intranuclear cascade step in one (randomly chosen) 
of the spectators.

Due to relativistic time dilatation secondaries
with high energies in the rest frame of the considered nucleus are
mostly formed outside of the spectator part of this nucleus whereas
those with low energies are formed inside. The latter may penetrate the
spectator and initiate intranuclear cascade processes. 
We note that this scheme is not Lorentz invariant. However, the
dependence of the results on the choice of the particular Lorentz frame
is assumed to be small. Therefore we avoid a generally possible but
technically cumbersome Lorentz invariant formulation of the intranuclear
cascade within the present approach.
Elastic and inelastic interactions with spectator nucleons are treated 
using the MC-model {\sc hadrin}~\cite{Haenssgen86}. This code
is based on measured cross sections and interaction channels up to a
laboratory momentum of 5~GeV. We apply {\sc hadrin} to
hadron-nucleon interactions up to 9~GeV and neglect those at higher
energies. Reinteractions beyond 5~GeV occur much less frequently than 
reinteractions below 5~GeV and a more detailed treatment would not
change the results discussed in this paper. Furthermore,
the absorption of low-energy mesons and antiprotons by interactions with
two-nucleon systems and Pauli's principle are taken into 
account~\cite{Moehring91}. If there was no interaction possible in the 
considered spectator we proceed with sampling a cascade step in the other
spectator. 

For secondaries produced in intranuclear cascade processes we 
apply the same formalism, i.e. a formation time is sampled, the secondary 
is transported to the end of the formation zone and reinteractions are
treated if they are possible. 
Due to these intranuclear cascade processes nucleons are knocked 
out of the residual spectator nuclei if their energy is high enough to
escape from the nuclear potential which will be discussed in the
following section.
\subsection{The calculation of nuclear excitation energies}
Excitation energies of projectile and target prefragments in
nucleus-nucleus collisions are calculated in a way which is similar to 
the one described in~\cite{Ferrari95a} for hadron-nucleus collisions. 

We assume that nucleons of a nucleus of mass $A$ and charge $Z$ move in
an effective nuclear potential defined by a sum of a Fermi-potential 
and a binding energy contribution
\begin{equation}
\label{nucpot}
V_i(A,Z)=\frac{\left[p_i^{\mbox{\scriptsize F}}(A,Z)\right]^2}{2m_i}+
E_i^{\mbox{\scriptsize bind}}(A,Z), \qquad i={\mbox{\rm proton, neutron}}
\end{equation}
with
\begin{eqnarray}
\label{nucpot1}
p_i^{\mbox{\scriptsize F}}(A,Z)&=&
\alpha_{\mbox{\scriptsize mod}}^{\mbox{\scriptsize F}}
\left[\frac{3h^3}{8\pi}\frac{N_i}{V_{\mbox{\scriptsize
A}}}\right]^{\frac{1}{3}}\nonumber\\
E_p^{\mbox{\scriptsize bind}}(A,Z)&=&E_{\mbox{\scriptsize bind}}(A,Z)-
E_{\mbox{\scriptsize bind}}(A-1,Z-1)\nonumber\\
E_n^{\mbox{\scriptsize bind}}(A,Z)&=&E_{\mbox{\scriptsize bind}}(A,Z)-
E_{\mbox{\scriptsize bind}}(A-1,Z)
\end{eqnarray}
and $N_p=Z$, $N_n=A-Z$.
$V_{\mbox{\scriptsize A}}$ stands for the volume of the corresponding
nucleus with an approximate nuclear radius $R_A=r_0A^{1/3}, r_0=1.29$~fm,
and $E_{\mbox{\scriptsize bind}}$ for its binding energy.
Modifications of the actual nucleon momentum distribution arising, for
instance, from nuclear skin effects are taken into account by the
reduction factor $\alpha_{\mbox{\scriptsize mod}}^{\mbox{\scriptsize
F}}$. This factor is adjusted by comparing the results of
the model to measured black particle production (see Section~\ref{ngnbcmp}) 
and is fixed at 0.5\footnote{In~\cite{Ferrari95a} we used
$\alpha_{\mbox{\scriptsize mod}}^{\mbox{\scriptsize F}}=0.75$ together
with $\tau_0=2$~fm/$c$, which were obtained comparing the model to h-A
data only. The parameters reported in this paper are based on a larger
data set, i.e. are based on both, data from hadron-nucleus and
nucleus-nucleus collisions}$^)$.
To be able to escape, positively charged particles have to penetrate
a Coulomb-barrier given by
\begin{equation}
\label{coulpot}
V_{\mbox{\scriptsize C}}(A,Z) = \frac{e^2}{4\pi\epsilon_0 r_0}
\frac{Z}{(1+A^{1/3})}.
\end{equation}
$e$ is the elementary charge and $r_0=1.29$~fm.

Due to the primary and secondary nucleon-nucleon interactions and to the
intranuclear cascade processes the colliding nuclei lose part of their
nucleons. These nucleons and the secondary hadrons which were formed
inside one of the nuclei have to escape from the effective nuclear 
potential.
Within our simplified approach the precise nuclear potentials of the
spectators during or after the primary or secondary
nucleon-nucleon interactions cannot be predicted. For baryons\footnote{
The fraction of baryons other than nucleons which are
created inside of one of the nuclei is small. They are assumed to move in 
a nucleon potential} we approximate it by
(\ref{nucpot}) and (\ref{nucpot1}) at mass $A_{\mbox{\scriptsize res}}$ and 
charge $Z_{\mbox{\scriptsize res}}$ of the corresponding nuclear spectator 
(prefragment) replacing in (\ref{nucpot1}) the modification factor 
$\alpha_{\mbox{\scriptsize mod}}^{\mbox{\scriptsize F}}$ by 
$\tilde{\alpha}_{\mbox{\scriptsize mod}}^{\mbox{\scriptsize F}}=
\alpha_{\mbox{\scriptsize mod}}^{\mbox{\scriptsize F}}+0.1$.
In case of mesons we apply an effective potential of 0.002~GeV. 
The escape of a hadron from the nuclear potential of a
prefragment implies that it transfers a recoil momentum to this
prefragment. To be definite, this means, e.g. for a nucleon
``leaving'' the projectile prefragment, that the energy of this nucleon 
in the projectile rest system is reduced by the effective nuclear potential
and, in case of a proton, has to be sufficiently high to overcome the 
Coulomb-barrier (\ref{coulpot}).
Its 3-momentum is rescaled correspondingly. Energy-momentum
conservation requires a recoil momentum transferred to the prefragment
which is equal to this correction.
Adding up these recoil momenta separately for the projectile and target
spectator, $(E_{\mbox{\scriptsize rcl}},\vec{p}_{\mbox{\scriptsize
rcl}})$, one obtains the 4-momenta of the prefragments
\begin{equation}
(E_{\mbox{\scriptsize res}},\vec{p}_{\mbox{\scriptsize res}})=
(M_{\mbox{\scriptsize A}},\vec{0})-\sum_{i=1}^{N_w} (E_i^{\mbox{\scriptsize
F}},\vec{p_i}^{\mbox{\scriptsize F}})+(E_{\mbox{\scriptsize
rcl}},\vec{p}_{\mbox{\scriptsize rcl}}).
\end{equation}
$N_w$ is the number of wounded nucleons in the corresponding nucleus
and $M_{\mbox{\scriptsize A}}$ is the mass of this nucleus.
For simplicity we omit indices referring the quantities to projectile and
target.

The excitation energy $U$ of a prefragment is defined as the energy
above the ground state mass $E_0$
\begin{eqnarray}
\label{excener}
&U&=E_{\mbox{\scriptsize res}}-E_0(A_{\mbox{\scriptsize
res}},Z_{\mbox{\scriptsize res}}) \nonumber\\
&E_0(A_{\mbox{\scriptsize res}},Z_{\mbox{\scriptsize res}})&=
Z_{\mbox{\scriptsize res}}m_{\mbox{\scriptsize p}}+
(A_{\mbox{\scriptsize res}}-Z_{\mbox{\scriptsize
res}})m_{\mbox{\scriptsize n}}
-E_{\mbox{\scriptsize bind}}
(A_{\mbox{\scriptsize res}},Z_{\mbox{\scriptsize res}}).
\end{eqnarray}
with $E_{\mbox{\scriptsize bind}}$ being the
corresponding binding energy which is either the experimentally
determined excess mass or is obtained from mass formulae for nuclides far
from the stable region. 
As examples, in Table~\ref{AAextab} we give average excitation energies 
and excitation energies per nucleon of target prefragments
after interactions of oxygen and aluminum with gold nuclei at different
energies. In asymmetric nucleus-nucleus interactions the excitation 
energies of the prefragments of the heavier target nucleus are slowly 
increasing with the beam energy and, at fixed energy, with the mass 
number of the projectile.
%
%
\section{$\gamma$-deexcitation of residual spectator nuclei}
%
%
At the end of the intranuclear cascade the prefragments are supposed to
be left in an equilibrium state, characterized by their masses, charges,
and excitation energies with no further memory of the steps which led to
their formation. Since the excitation energies can be higher than the
separation energies, nucleons and light fragments ($\alpha$,d,$^3$H,
$^3$He) can still be emitted. A detailed account on the evaporation
treatment which follows the intranuclear 
cascade was given in~\cite{Ferrari95a}. Furthermore, a model for high
energy fission and a Fermi Break-up model for light nuclei where an
excited prefragment is supposed to disassemble just in one step into two 
or more fragments were discussed in~\cite{Ferrari95a}.

The evaporation stage ends when the nuclear excitation energy becomes 
lower 
than all separation energies for nucleons and fragments. This residual 
excitation energy is then dissipated through emission of photons. 
In reality, photon emission occurs even during the preequilibrium and 
evaporation stages, in competition with particle emission, but its
relative probability is low, and it is presently neglected in the model.

$\gamma$-deexcitation proceeds through a cascade of 
consecutive photon emissions, until the ground state is reached. The 
cascade is assumed to be statistical as long as the excitation energy 
is high enough to allow the definition of a continuous nuclear level 
density. Below a (somewhat arbitrary) threshold, set at the pairing gap 
value, the cascade goes through transitions between discrete levels. 

The statistical model formulation for the $\gamma$-ray emission probability
is again similar to those for evaporation and 
fission~\cite{Bartholomew75,Bergqvist70}
\begin{equation}
P(E_\gamma)dE_\gamma = \frac{\rho_f(U_f)}{\rho_i(U_i)}\sum_L{f(E_\gamma,L)}
dE\gamma
\label{eq:initg}
\end{equation}
where $L$ is the multipolarity of the $\gamma$ transition. 
The strength functions 
$f(E,L)$ can be either derived from photoabsorption cross sections or 
calculated from single-particle estimates of transition strengths. 
The former approach is more sophisticated but requires the knowledge of the 
resonance parameters for all isotopes; 
the latter approach is easier and sufficient for 
a first order  estimate of the $\gamma$ spectral distribution. We thus 
assume
\begin{equation}
f(E_\gamma,L) = c_L \cdot F_L(A) \cdot E_\gamma^{(2L+1)}
\end{equation}
where $E_\gamma^{(2L+1)}$ is the energy dependence for multipolarity
$L$. For the $c_L$ coefficients we adopted the Weisskopf single particle
estimates~\cite{Wilkinson60}. The $F_L(A)$ factors
have been included to partially account for the many effects that 
bring to deviations from the single particle estimates. They are
rough A dependent averages of the  hindrance and
enhancement factors given in~\cite{Endt81}. 
Only E1, M1, and E2 transitions have been considered.
The assumed level density $\rho$ is the same as in the evaporation 
part described in~\cite{Ferrari95a}, but the ratio  of exponentials 
coming from the level densities has been approximated as first order 
expansion around $E_\gamma=0$. This is equivalent to the 
assumption of a constant nuclear temperature at low excitation energies, 
which is often used in the analysis of photon emission following neutron 
capture~\cite{Bartholomew75,Dilg73}.

As a result, one obtains the expressions for the emission probabilities for
the considered multipoles. Since competition between photon and 
particle emission  is neglected at the present status  of the model,
only the relative values are of interest
\begin{eqnarray}
\label{eq:pegam}
P(L,E_\gamma)dE_\gamma &= &\tilde C_L 
E_\gamma^{(2L+1)}e^{\frac{E_\gamma}{T}}dE_\gamma \\
\frac{\tilde C_{M1}}{\tilde C_{E1}}&
=&0.31A^{-\frac{2}{3}}\frac{F_{M1}(A)}{F_{E1}(A)} \nonumber \\
\frac{\tilde C_{E2}}{\tilde C_{E1}}&
=&7.2\cdot 10^{-7}A^{\frac{2}{3}}\frac{F_{E2}(A)}{F_{E1}(A)} \ \ 
{\mbox{\rm MeV}}^{-2}
\nonumber
\end{eqnarray}
$T$ is the nuclear temperature at the initial excitation energy $U$, 
taken as $U-\Delta = a T^2$, $a$ being the usual level density constant
(see Eq.15 of~\cite{Ferrari95a}) and $\Delta $ the pairing energy. 

A first sampling is performed on the integrated photon emission 
probabilities 
to choose the character( electric or magnetic) and the multipole order 
of the emitted photon, and a second 
sampling is performed to determine the emission energy according 
to the selected multipolarity. For  both steps 
the full energy range $0\le E_\gamma\le U$ is used, even though the 
intrinsic limit of validity would be $0\le E_\gamma\le (U-\Delta)$. 
After emission, all parameters are updated on the basis of the new 
excitation energy, and another statistical emission is performed, until
the excitation energy falls below the preset "discrete level threshold".
This threshold  has been set to the pairing energy for even-even or odd 
mass nuclei; for odd-odd nuclei the threshold corresponds to the (known 
or approximated) first excited level.

For many isotopes the experimentally determined values of the first and 
second excited levels have been tabulated in the code, for the others a 
rotational-like structure is assumed, with level energy given by :
\begin{equation}
U_I=\frac{\hbar^2}{2\cal I}I(I+1)
\end{equation} 
where $I$ is the level spin (integer for even-mass , half integer for 
odd mass nuclei, 0 or 1/2 for the ground state), 
and $\cal I$ is the nuclear moment of 
inertia, taken as  0.4 times that of a  rigid body . The last steps of 
the $\gamma$ cascade consist of $\delta I$=2 transitions among these 
rotational levels, down to the ground state. When known levels are 
tabulated, the cascade is forced to pass through them.
All photons are  emitted isotropically, since from the evaporation stage 
we have no information on the residual nucleus spin and polarization. 

The $\gamma$ deexcitation model has been developed and tested within the 
cascade-preequilibrium-evaporation 
model {\sc peanut}~\cite{Sala94,Fasso94b}, and it has been 
easily coupled to {\sc dtunuc} and {\sc dpmjet}
since its algorithm depends only on the residual 
nucleus mass, Z and excitation energy after evaporation, not on the 
details of the preceding interaction history.
Two examples of its performances are shown in Figs. \ref{tigamma} and
\ref{wgamma}. Both, the total photon multiplicity and the shape
of the spectra are well reproduced. 
%
%
\section{Comparing target-associated particle production to experimental
         data}
\subsection{\label{ngnbcmp}Grey and black particles and correlations}
As discussed in Sect.\ref{fzcAA} nucleons might be knocked out by
intranuclear cascade processes if their energies are sufficiently high to
escape from the nuclear potential. The kinetic energies are typically of
the order of 20 to 400~MeV in the rest frame of the considered nucleus.
In contrast, nucleons and light fragments evaporated from the excited
prefragment possess less energy. This allows to study both mechanisms of
low energy particle production rather independently, apart from the
fact that in our model the cascade particles contribute to the excitation 
of the prefragment and, therefore, indirectly to the multiplicity of
the evaporated particles.

There are only a few experiments known to us in which 
target associated particle production in interactions with nuclear
projectiles is investigated. The most detailed results were obtained in
experiments using emulsion targets. The emulsions usually consist of a
component of light nuclei (H,C,N,O) and a component of heavy nuclei
(Ag,Br). The appearance of slow charged particles and light fragments in
these experiments has led to their subdivision into ``grey'' and
``black'' particles. 
Grey and black particles roughly coincide
in energy with the above mentioned cascade and evaporation particles.

In Table~\ref{AAmultngnbtab} we give average values for multiplicities of
grey and black particles in interactions of oxygen, silicon, sulfur, and
gold nuclei with emulsions at different beam energies as calculated 
within our model.
The experimental values are results of the
EMU01-Collab.~\cite{Adamovich91,Adamovich95c}. Our results are 
superpositions of
multiplicities obtained with hydrogen, light and heavy nuclear targets
weighted with the fractions 0.13, 0.31, and 0.56 resp., corresponding to
the emulsion composition used in the experiments~\cite{Adamovich88}. 
Furthermore, we apply the definitions for ``grey'' and ``black'' as 
given by the EMU01-Collab. Grey particles are protons, pions, and kaons
with kinetic energies between 26 and 375 MeV, 12 and 56 MeV, and 20 and
198 MeV, resp. Black particles are singly and multiply charged having 
lower
energies than the grey particles. Apart from the results for $\langle
N_g\rangle$ at low energies where the model seems to overestimate the 
measured multiplicities the agreement is satisfactory. Slow protons
contribute about 79\% and slow charged pions about 19\% to the
calculated grey particle multiplicity in S-Emulsion interactions at 
200~GeV/nucleon.

In Figs.\ref{Aemung}--\ref{auemunb} we show multiplicity distributions
for grey and black particles in oxygen- and sulfur-emulsion interactions 
at 200~GeV/nucleon and in gold-emulsion interactions at 11.6~GeV/nucleon.
They are compared to data of the 
EMU01-Collab.~\cite{Adamovich95c,Adamovich95a} and, for oxygen and
sulfur projectiles, in addition to data of the 
KLM-Collab.~\cite{Dabrowska93b}. 
The contribution of the light emulsion component dominates the shapes of
the distributions at low multiplicities, whereas for $\langle N_{g,b}
\rangle \stackrel{>}{\sim} 5$ the distributions are determined by 
interactions with the
heavy target nuclei, silver and bromine. Again, within our calculations 
we have
applied the above given definitions for ``grey'' and ``black''. The
differences between the experimental data of both Collaborations can be
possibly explained by the cut of 0.25 between the
Lorentz-$\beta$ values of grey and black particles used by the
KLM-Collab.~\cite{Dabrowska93b}. For the oxygen- and sulfur-emulsion
interactions (Figs.\ref{Aemung},\ref{Aemunb}) our results agree with the
measured distributions apart from a possible underestimation of the tail
of the $N_g$-distributions at high multiplicities.
In case of gold projectiles the calculated grey particle distribution
(Fig.\ref{auemung}) shows for $N_g>10$ a rather flat shape, which is
different from the measured distribution. This indicates more cascading 
and probably bigger target prefragments than seen in the experiment.
This underlines the
fact that our model of intranuclear cascading and excitation energy
calculation is too simplified to understand slow particle production in
very asymmetric nucleus-nucleus collisions in the rest system of the
smaller nucleus. This nucleus gets almost completely disintegrated
already in the primary and secondary interactions and a
multifragmentation model could be more appropriate for the description 
of the nuclear disintegration processes.

The reasonable description of the multiplicity distributions in 
oxygen- and sulfur-emulsion interactions suggests that the model should 
be able to 
reproduce measured correlations between grey, black, and shower particle 
multiplicities. In Figs.\ref{semuninscorr}a,b we show the average number
of grey and black particles depending on the number of shower particles.
The agreement with the data~\cite{Adamovich95a} is reasonable except for 
the average number of black particles around $N_s\approx 200$.
The correlations between heavy (=grey+black) and shower
particle multiplicities are compared to data~\cite{Adamovich95a} in
Figs.\ref{semuninscorr}c,d. Our model is able to reproduce the measured
correlations within their statistical uncertainties. Finally, in
Figs.\ref{semungnbcorr}a,b we give the correlations between grey and
black particle multiplicities. We obtain slightly less grey 
particles in high-multiplicity S-Ag/Br interactions than seen in the
experiment.

In order to investigate the kinematics for slow particle production
angular distributions of grey and black particles are compared to
data on oxygen- (Figs.\ref{Aemunith}a,c) and sulfur-emulsion
interactions (Figs.\ref{Aemunith}b,d) at
200~GeV/$c$/nucleon~\cite{Adamovich91,Dabrowska93b}. Whereas in the
forward direction protons dominate the angular distributions of grey
particles, in the backward direction pions produced in intranuclear 
cascade
processes become important. Parametrizing the angular distributions of 
grey particles with $f(\cos\Theta_g)=K\exp(b\cos\Theta_g)$ we obtain
$K=0.39$ and $b=1.25$. The experimental values are $K=0.44$ and
$b=0.92$, i.e. our results show a bigger slope.
The angular distributions of black particles are
in good agreement with the data which supports the fact that we
correctly describe the recoil momentum transferred to the residual
nucleus.

For a more detailed investigation of slow particle production it is
needed to compare the MC-results to data obtained in experiments which
(i) are using one kind of target nucleus rather than a mixture as the
emulsions are and (ii) which are able to identify particles and may
therefore separate slow protons and light fragments from pions. Such an
experiment with oxygen projectile nuclei and several target nuclei 
was performed by the WA80-Collab. and results were published
in~\cite{Albrecht93b}. These data are subject to the limited detector
acceptance~\cite{Albrecht93b,Baden82,Gutbrod96pc} and any comparisons
of model-results which were not filtered through the acceptance
functions have to be taken with care. The following cuts have been
applied to our results in order to be as close to the experimental
conditions as possible~\cite{Albrecht93b}: (i) an event is
considered if the energy within a cone of 0.3$^{\circ}$ is less than
88\% of the beam energy and (ii) only protons and singly charged
fragments having zenit angles between 60$^{\circ}$ and 160$^{\circ}$ and
kinetic energies between 30~MeV/nucleon and 400~MeV/nucleon were taken
into consideration. In Fig.\ref{oAwa80} we show the multiplicity 
distributions of slow fragments from oxygen-nucleus interactions at
200~GeV/$c$/nucleon together with the WA80-data. Our results are 
normalized to the data.
\subsection{\label{Arescmp}Residual target nuclei}
The heavy prefragments which are left after the evaporation of nucleons 
and light nuclei can be considered as heavy fragments produced in a
spallation or deep spallation process. The high energy fission and Fermi
Break-up models contribute to a further modification of the mass yield
if these processes are physically possible. Since fragmentation of
prefragments with masses above $A_{\mbox{\scriptsize res}}=18$ is not 
treated we cannot expect to agree with measured mass yields in 
nucleus-nucleus collisions. 
However, as it is shown in Fig.\ref{cag22resA} for carbon-silver
interactions at 25.2~GeV/nucleon our cross sections for the production
of residual target nuclei differ by not more than a factor of two from the
measured values~\cite{Porile79} within the mass range between 30 and 100. 
The rising
yields of residual nuclei close to the target mass ($A_T-5\le
A_{\mbox{\scriptsize res}}\le A_{\mbox{\scriptsize T}}$) are not described 
within our model
since we do not consider such processes like quasi-elastic scattering
and treat the nuclear potential only in a rough manner (c.f. discussion
of target fragmentation in proton-nucleus interactions 
in~\cite{Ferrari95a}). The calculated yields at $A_{\mbox{\scriptsize
res}}=$2,3,4 represent nuclei evaporated from the prefragments.
%
%
\section{Applications}
\subsection{Residual nuclei and high-energy photons from interactions 
            of cosmic-ray nuclei in the atmosphere}
For studying cosmic ray cascades the forward region is of main
importance. In case of nucleus-air interactions this involves the
projectile associated particle production and the fragmentation and
deexcitation of the projectile prefragment. Since oxygen and iron nuclei
are typical cosmic ray nuclei we restrict the following
study to oxygen- and iron-air interactions. Of course, the interactions
of all cosmic ray nuclei will be treated by our event generators, if
they are used in a cosmic ray cascade code.

In order to investigate to which extent the results discussed in this
paper depend on the MC-realization of the DPM we compare
multiplicities obtained with {\sc dpmjet-ii} and
{\sc dtunuc~2.0}. In Tables~\ref{feairmultab} and \ref{oairmultab}
we present shower, grey, and black particle multiplicities (the
definitions are the same as for the comparisons with the emulsion data
in Section~\ref{ngnbcmp}) in iron- and oxygen-air interactions at 
different
laboratory momenta of the projectile. Furthermore, the average
multiplicities of evaporation products and deexcitation photons 
from the projectile and target fragments are given. Since {\sc dtunuc~2.0}
cannot yet be applied to interactions at nucleon-nucleon c.m. energies
beyond 1~TeV ($p_{\mbox{\scriptsize Lab}}\approx$533~TeV/$c$)
multiplicities obtained with this event generator are
presented only for the two lower momenta, 0.2A~TeV/$c$ and 20A~TeV/$c$
(second lines). It can be concluded that, apart from the shower particle
multiplicities, differences in the given multiplicities caused by the 
particular MC-realization of the DPM are smaller than the statistical 
uncertainties of the calculations.

The mass yields of residual projectile nuclei in oxygen-air interactions
are shown for two different energies
in Fig.\ref{AairAres}a. We observe that the yield of nuclei is
shifted towards smaller mass numbers as the oxygen energy increases.
Light nuclei ($A_{\mbox{\scriptsize res}}\le 4$) clearly dominate the
residual nuclei with higher masses. Note, that the prefragments obtained
after the intranuclear cascade were subject to both, the fragmentation
by the evaporation and the Fermi Break-up models. Structures apparent in
the mass yields are not due to statistical fluctuations but result from
the fragmentation process. In Fig.\ref{AairAres}b we present the mass
yields of residual projectile nuclei in iron-air interactions.
We obtain higher yields of small
nuclei with increasing energy whereas the shapes of the yields at the
two energies are similar above $A_{\mbox{\scriptsize res}}=25$. We
stress again that for $A_{\mbox{\scriptsize res}}>16$ apart from
evaporation of light fragments no further fragmentation is treated. The
yields close to the iron mass might be underestimated as it is discussed
in Sect.\ref{Arescmp}. 

Within our MC-event generators there are two main mechanisms of photon
production, the decay processes of hadrons and the nuclear deexcitation
of the prefragments. In order to investigate to which extent they
contribute to the total photon spectrum we show in Figs.\ref{feairgpr}a,b
the pseudorapidity distributions of photons from $\pi^0$-decay and from
deexcitation processes separately. It is obvious that photon production
by nuclear deexcitation contributes only a minor fraction to the total
photon yield in forward and backward direction. Integrating the two
peaks in the distributions of the deexcitation photons separately one
obtains the average multiplicity of photons from the deexcitation of the
projectile and target prefragments, which is directly correlated to 
their average excitation energies after the evaporation step. 
It demonstrates, that the average residual
excitation energy remains about constant with the collision energy above
200~GeV/nucleon and that the average residual excitation energy of the
prefragments originating from the iron nuclei is significantly higher as 
compared to the one from air nuclei.

Finally, in Figs.\ref{feairgxf}a,b we show the Feynman-$x_F$
distributions of photons in the nucleon-nucleon c.m. frame at a
laboratory energy of 200000~TeV/nucleon. Again, the contribution of
deexcitation photons can be neglected in comparison to the decay
photons.
\subsection{Residual nuclei production at heavy-ion colliders}
As for cosmic ray cascades a realistic description of projectile and
target associated particle production and of the production of residual
nuclei becomes important if one is investigating particle production in
the forward or backward fragmentation region. Furthermore, in detector
simulations for heavy ion experiments, such as simulations of the
performances of trigger Zero Degree Calorimeters, as well as in 
radiation shielding
calculations for existing or planned heavy ion colliders a model for the
production of residual nuclei is an essential ingredient for the
estimation of particle cascades in matter initiated by nuclei.

In Fig.\ref{pbpbAres} we give cross sections for residual nuclei
produced in lead-lead collisions at a RHIC- c.m. energy of
200~GeV/nucleon. The nuclei with masses up to $A_{\mbox{\scriptsize
res}}=4$ are evaporation products. Apart from Fermi Break-up for the few
light fragments no further fragmentation was performed, i.e. the
cross sections shown are production cross sections for residual nuclei
from one of the lead nuclei after the evaporation step.

Again, as it was shown in Sect.\ref{ngnbcmp} the models describe
slow particle production in asymmetric collisions rather satisfactorily  
but it has still to be proven that they can equally well be applied to 
collisions of identical or nearly identical heavy nuclei.
%
%
\section{Summary and conclusions}
In the present paper, models for formation zone intranuclear cascades in 
spectator prefragments, the evaporation of nucleons and light fragments,
high energy fission processes, and the fragmentation of light nuclei were 
discussed and applied to peripheral nucleus-nucleus collisions.
Furthermore, we have introduced a model for nuclear deexcitation by photon 
emission.

Comparing our results to data on slow particle and fragment production
it is shown that together with the two-component Dual Parton Model for
nucleus-nucleus collisions main features of projectile and target
associated particle production are well described. In particular,
results of calculations with {\sc dpmjet-ii} and {\sc dtunuc 2.0} were
found to be in agreement with multiplicity and angular distributions of grey 
and black prongs and with correlations between grey, black, and shower
particle multiplicities. 
Although we stress, that the model is only valid for peripheral
collisions, we use it actually for minimum bias nucleus-nucleus collisions,
where a small fraction of central collisions contributes, for which the
model cannot be expected to be reliable. Therefore, the model results 
might deviate from 
data on multiplicity distributions of grey and black prongs and slow
fragments at the high multiplicity tails and we might expect that the model 
misses some part of the residual nucleus and nuclear fragment distribution.

The $\gamma$-deexcitation model is compared to low energy data. It is
discussed that with rising collision energy, the contribution from 
deexcitation photons becomes relatively unimportant as compared to photons 
from hadronic decays.

However, even in the absence of a multifragmentation model the
production of residual nuclei turned out to be an acceptable
approximation for fragment production within a MC-event generator to be
used to describe hadronic interactions in cascade codes like 
{\sc fluka}~\cite{Fasso94b,Fasso94a,Aarnio94} or 
{\sc hemas}~\cite{Battistoni95}.
Using the discussed models within the simulation of cosmic ray cascades 
allows to go beyond the ``superposition model''~\cite{Forti90} and similar 
approximations for the collisions of cosmic ray nuclei in the atmosphere. 
Likewise, in heavy ion experiments, the nuclear fragments and residual
nuclei are often very important to simulate the performance of components
of the detectors in extreme forward or backward direction, like the Zero
Degree Calorimeters of some existing experiments.
%
%
\section*{Acknowledgements}
One of the authors (S.R.) is grateful to F.W.\ Bopp and R.\ Engel for 
stimulating discussions. S.R. acknowledges the collaboration with 
R.\ Engel on the link between the {\sc dtunuc} and {\sc phojet} event 
generators.
We thank H.\ Gutbrod, K.-H.\ Kampert, and E.\ Stenlund from the 
WA80-Collaboration for helpful comments on the EMU01- and WA80-data.
%
%
\clearpage

%
%
\clearpage
\section*{Tables}
\begin{table}[htb]
\caption{\label{AAextab} 
      Average excitation energies and excitation energies per
      nucleon of target prefragments produced in oxygen- and 
      aluminum-gold interactions before evaporation
      are given for different momenta per nucleon of the projectile
      nucleus.
        }
\medskip
\begin{center}
\renewcommand{\arraystretch}{1.5}
\begin{tabular}{|c||c|c||c|c|} 
\hline
&\multicolumn{2}{|c||}{O-Au}&\multicolumn{2}{|c|}{Al-Au}\\
$p_{\mbox{\scriptsize Lab}}$&
$\langle U\rangle$& $\langle \frac{U}{A_{\mbox{\scriptsize res}}}\rangle$&
$\langle U\rangle$& $\langle \frac{U}{A_{\mbox{\scriptsize res}}}\rangle$\\
(A~GeV/$c$)& (MeV) & (MeV) & (MeV) & (MeV) \\
\hline \hline
  20  &  592.0 & 5.6 & 550.0 & 5.9 \\ \hline
  30  &  554.0 & 5.5 & 537.0 & 6.0 \\ \hline
  50  &  567.0 & 5.6 & 546.0 & 6.2 \\ \hline
 100  &  567.0 & 5.7 & 573.0 & 6.7 \\ \hline
 200  &  573.0 & 5.8 & 591.0 & 7.1 \\ \hline
 500  &  603.0 & 6.4 & 606.0 & 7.1 \\ \hline
1000  &  619.0 & 6.6 & 621.0 & 7.3 \\ \hline
2000  &  653.0 & 6.7 & 652.0 & 7.7 \\ \hline
\end{tabular}
\end{center}
\end{table}
\begin{table}[htb]
\caption{\label{AAmultngnbtab}
    Multiplicities of grey ($N_g$) and black ($N_b$) particles 
    in interactions of nuclei with emulsions are given for different 
    projectile nuclei and for different projectile energies.
    The experimental data from the EMU01-Collab. were taken
    from~\protect\cite{Adamovich91} if not explicitly indicated in 
    the table.}
\medskip
\begin{center}
\renewcommand{\arraystretch}{1.5}
\begin{tabular}{|c|c||c|c|c|c|} \hline
&$E_{\mbox{\scriptsize Lab}}$&
\multicolumn{2}{|c|}{$\langle N_g \rangle$}&
\multicolumn{2}{|c|}{$\langle N_b \rangle$} \\
& (GeV/nucleon)& DTUNUC& Exp.& DTUNUC& Exp.
\\ \hline \hline
O-Emul. & 14.6 &6.1 & 5.2$\pm$0.2  &4.7 & 4.8$\pm$0.2  \\ \hline
        &  60  &4.3 & 5.7$\pm$0.4  &4.4 & 4.5$\pm$0.2  \\ \hline
        & 200  &4.2 & 4.3$\pm$0.3  &4.3 & 4.1$\pm$0.2  \\ \hline
Si-Emul.& 14.6 &7.4 & 5.4$\pm$0.3  &4.6 & 4.6$\pm$0.2  \\ \hline
S-Emul. & 200  &4.4 & 4.7$\pm$0.3  &4.2 & 3.9$\pm$0.2  \\ \hline
Au-Emul.& 11.6 &9.7 & 5.8$\pm$0.19 \protect\cite{Adamovich95c}
&3.6 & 3.52$\pm$0.12 \protect\cite{Adamovich95c} \\ \hline
\end{tabular}
\end{center}
\end{table}
\begin{table}[htb]
\caption{\label{feairmultab} 
    Multiplicities of shower ($N_s$), grey ($N_g$), and black ($N_b$)
    particles in iron-air interactions are given for different
    momenta per nucleon in the laboratory frame. In addition,
    multiplicities of evaporated protons ($N_p$), neutrons ($N_n$), 
    and heavy fragments ($N_{hf}$) and deexcitation photons ($N_{\gamma}$) 
    are presented for the projectile (iron) and target (air=N) fragments.
    At 0.2A~TeV/$c$ and 20A~TeV/$c$ we give in the first line the 
    multiplicities calculated with {\sc dpmjet-ii} and in the second line
    those calculated with {\sc dtunuc 2.0}. At the two highest energies
    only {\sc dpmjet-ii} predictions are given.
        }
\medskip
\begin{center}
\renewcommand{\arraystretch}{1.5}
\begin{tabular}{|c||c|c|c||c|c|c|c||c|c|c|c|} \hline
$p_{\mbox{\scriptsize Lab}}$&
\multicolumn{3}{|c||}{}&\multicolumn{4}{|c||}{Projectile fragments}&
\multicolumn{4}{|c|}{Target fragments} \\
(A~TeV/$c$)& $\langle N_s\rangle$ & $\langle N_g\rangle$ 
& $\langle N_b\rangle$ 
& $\langle N_p\rangle$ & $\langle N_n\rangle$ 
& $\langle N_{\gamma}\rangle$ & $\langle N_{hf}\rangle$
& $\langle N_p\rangle$ & $\langle N_n\rangle$ 
& $\langle N_{\gamma}\rangle$ & $\langle N_{hf}\rangle$
\\ \hline \hline
0.2    & 82.8 & 1.3 & 1.7 & 2.5 & 4.0 & 1.9 & 2.5 
                          & 0.54& 0.62& 0.10& 1.2 \\ \hline
       & 78.1 & 1.4 & 1.7 & 2.2 & 4.1 & 2.0 & 2.4
                          & 0.51& 0.58& 0.10& 1.1 \\ \hline
20     & 256.5& 1.2 & 1.6 & 2.8 & 4.5 & 1.7 & 2.9 
                          & 0.57& 0.65& 0.08& 1.2 \\ \hline
       & 234.6& 1.4 & 1.6 & 2.7 & 4.6 & 1.8 & 2.8
                          & 0.55& 0.63& 0.08& 1.1 \\ \hline
2000   & 641.3& 1.1 & 1.5 & 3.2 & 4.9 & 1.4 & 3.2 
                          & 0.56& 0.65& 0.07& 1.1 \\ \hline
200000 &1260.9& 1.1 & 1.5 & 3.5 & 5.2 & 1.2 & 3.4 
                          & 0.58& 0.65& 0.07& 1.1 \\ \hline
\end{tabular}
\end{center}
\end{table}
\begin{table}[htb]
\caption{\label{oairmultab} 
    As in caption of Table~\protect\ref{feairmultab} but for oxygen-air
    interactions.
        }
\medskip
\begin{center}
\renewcommand{\arraystretch}{1.5}
\begin{tabular}{|c||c|c|c||c|c|c|c||c|c|c|c|} \hline
$p_{\mbox{\scriptsize Lab}}$&
\multicolumn{3}{|c||}{}&\multicolumn{4}{|c||}{Projectile fragments}&
\multicolumn{4}{|c|}{Target fragments} \\
(A~TeV/$c$)& $\langle N_s\rangle$ & $\langle N_g\rangle$ 
& $\langle N_b\rangle$ 
& $\langle N_p\rangle$ & $\langle N_n\rangle$ 
& $\langle N_{\gamma}\rangle$ & $\langle N_{hf}\rangle$
& $\langle N_p\rangle$ & $\langle N_n\rangle$ 
& $\langle N_{\gamma}\rangle$ & $\langle N_{hf}\rangle$
\\ \hline \hline
0.2    & 45.4 & 1.3 & 2.1 & 0.74& 0.80& 0.20& 1.8 
                          & 0.64& 0.70& 0.13& 1.5 \\ \hline
       & 43.9 & 1.4 & 2.0 & 0.73& 0.82& 0.19& 1.7
                          & 0.62& 0.68& 0.12& 1.4 \\ \hline
20     & 144.8& 1.3 & 1.9 & 0.85& 0.90& 0.17& 1.7 
                          & 0.68& 0.71& 0.11& 1.5 \\ \hline
       & 126.0& 1.4 & 2.0 & 0.80& 0.93& 0.16& 1.8
                          & 0.65& 0.76& 0.11& 1.4 \\ \hline
2000   & 360.6& 1.2 & 1.9 & 0.91& 0.95& 0.14& 1.7 
                          & 0.71& 0.80& 0.10& 1.4 \\ \hline
200000 & 751.6& 1.2 & 1.9 & 0.91& 0.98& 0.12& 1.6 
                          & 0.72& 0.79& 0.09& 1.4 \\ \hline
\end{tabular}
\end{center}
\end{table}
%
%
\clearpage
\section*{Figure Captions}
\begin{enumerate}
\item \label{tigamma} 
      Photon spectrum resulting from the reaction Ti(n,x) at 19 MeV. The
      dashed histogram represents {\sc peanut} results with errors. 
      Dots are experimental data from~\protect\cite{Morgan79}.
\item \label{wgamma} 
      As in Fig.\protect\ref{tigamma} for Tungsten. 
      Experimental points are from~\protect\cite{Dickens73}.
\item \label{Aemung} 
      Grey particle multiplicity distributions for interactions of
      oxygen (a) and sulfur (b) with emulsion nuclei
      are plotted together with experimental results of the 
      EMU01-Collab.~\protect\cite{Adamovich95a,Adamovich89} and of the
      KLM-Collab.~\protect\cite{Dabrowska93b}.
\item \label{auemung} 
      Grey particle multiplicity distributions in gold-emulsion
      interactions are shown in comparison with data of the
      EMU01-Collab.~\protect\cite{Adamovich95c}.
\item \label{Aemunb} 
      Black particle multiplicity distributions for interactions of
      oxygen (a) and sulfur (b) with emulsion nuclei
      are plotted together with experimental results of the 
      EMU01-Collab.~\protect\cite{Adamovich95a,Adamovich89} and of the
      KLM-Collab.~\protect\cite{Dabrowska93b}.
\item \label{auemunb} 
      Black particle multiplicity distributions in gold-emulsion
      interactions are shown in comparison with data of the
      EMU01-Collab.~\protect\cite{Adamovich95a}.
\item \label{semuninscorr} 
      Correlations between grey ($N_g$), black ($N_b$), and shower 
      ($N_s$) particle multiplicities in interactions 
      of sulfur with emulsion nuclei are compared to experimental 
      results~\protect\cite{Adamovich95a}.
\item \label{semungnbcorr} 
      The correlations between grey ($N_g$) and black ($N_b$) particle 
      multiplicities in sulfur-emulsion interactions are compared to 
      experimental results~\protect\cite{Adamovich95a}.
\item \label{Aemunith} 
      Angular distributions of grey (a,b) and black (c,d) particles
      in oxygen- and sulfur-emulsion interactions are shown together
      with data~\protect\cite{Adamovich91,Dabrowska93b}.
\item \label{oAwa80} 
      Multiplicity distributions of slow protons and singly charged
      fragments from oxygen-nucleus interactions are compared to results of 
      the WA80-Collab.~\cite{Albrecht93b}. See the text for kinematical
      cuts applied.
\item \label{cag22resA} 
      Mass distributions of residual target nuclei produced in carbon-silver 
      interactions at 25.2~GeV/nucleon as obtained with the
      model are compared to experimental results~\protect\cite{Porile79}.
\item \label{AairAres} 
      Cross sections for the production of residual projectile nuclei 
      in interactions of oxygen (a) and iron (b) nuclei with air
      (nitrogen) are shown for two different laboratory energies.
\item \label{feairgpr}
      Pseudorapidity distributions of photons produced in oxygen- (a) and
      iron-air (b) interactions at laboratory energies of 0.2~TeV/nucleon 
      (filled symbols) and of 2000~TeV/nucleon (open symbols) are given. 
      We show the contributions of photons resulting from decays and those 
      from deexcitation processes of the projectile prefragments separately.
      The distributions are plotted in the nucleon-nucleon c.m. frame.
\item \label{feairgxf}
      Feynman-$x_F$-distributions of photons produced in iron-air
      interactions at 200000 TeV/nucleon are shown. In (a) the contributions
      of photons from decays (stars) and from deexcitation processes of
      the projectile prefragments (line) as calculated within our model are
      presented. In (b) only deexcitation photons are plotted.
\item \label{pbpbAres}
      Cross sections for the production of residual nuclei
      in lead-lead collisions at an energy of 200~GeV in the
      nucleon-nucleon c.m. system. Shown is the fragmentation yield for
      one of the colliding nuclei.
\end{enumerate}
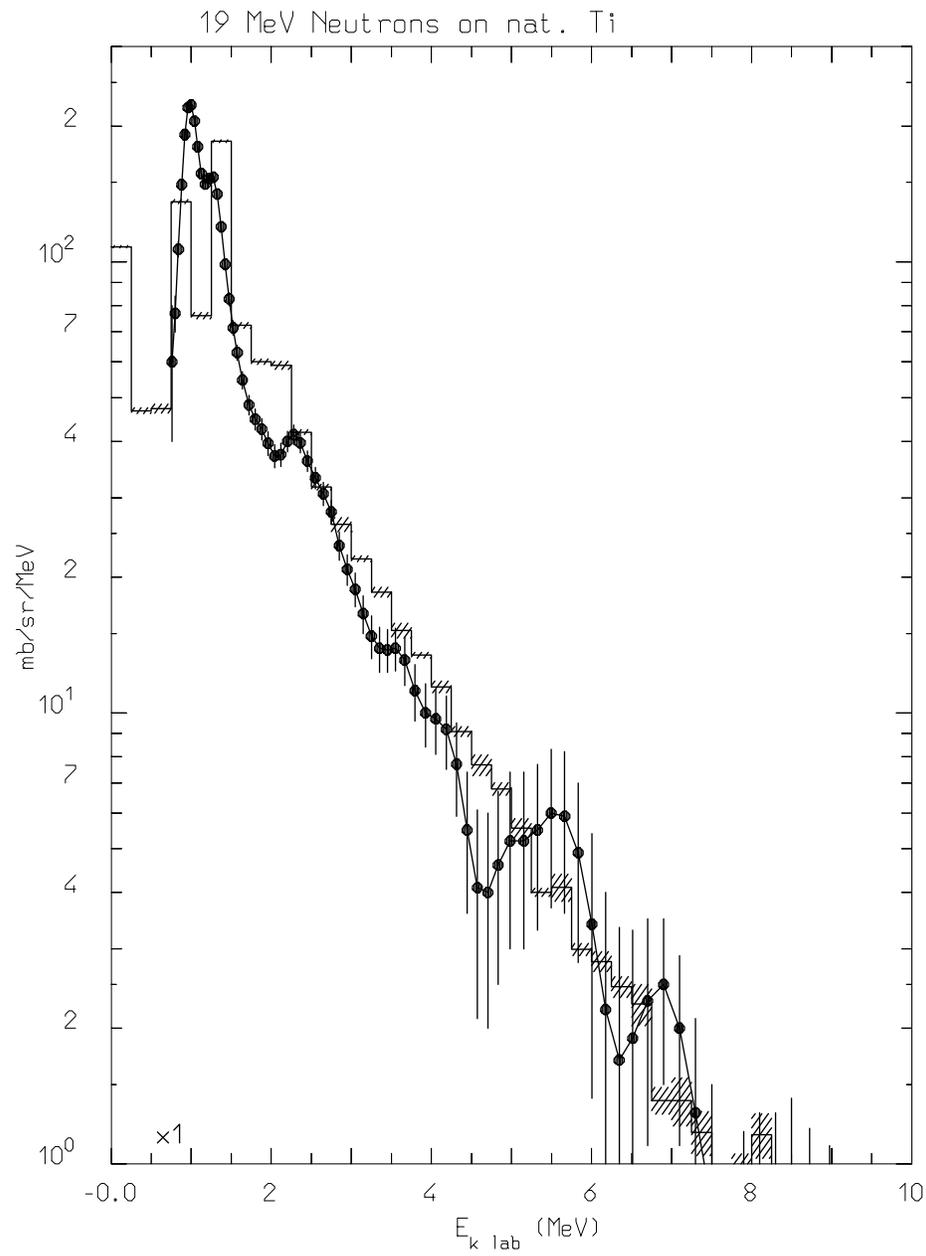
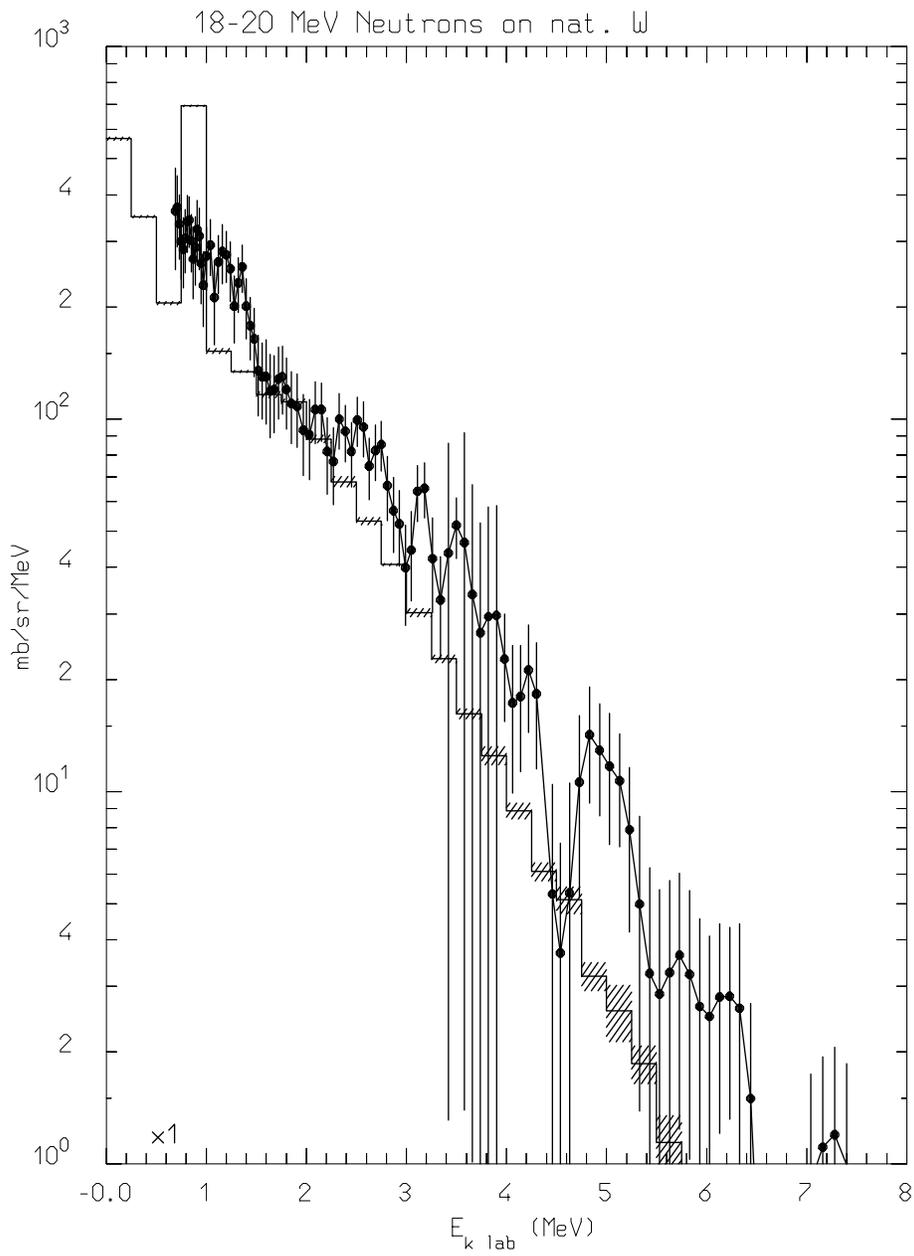
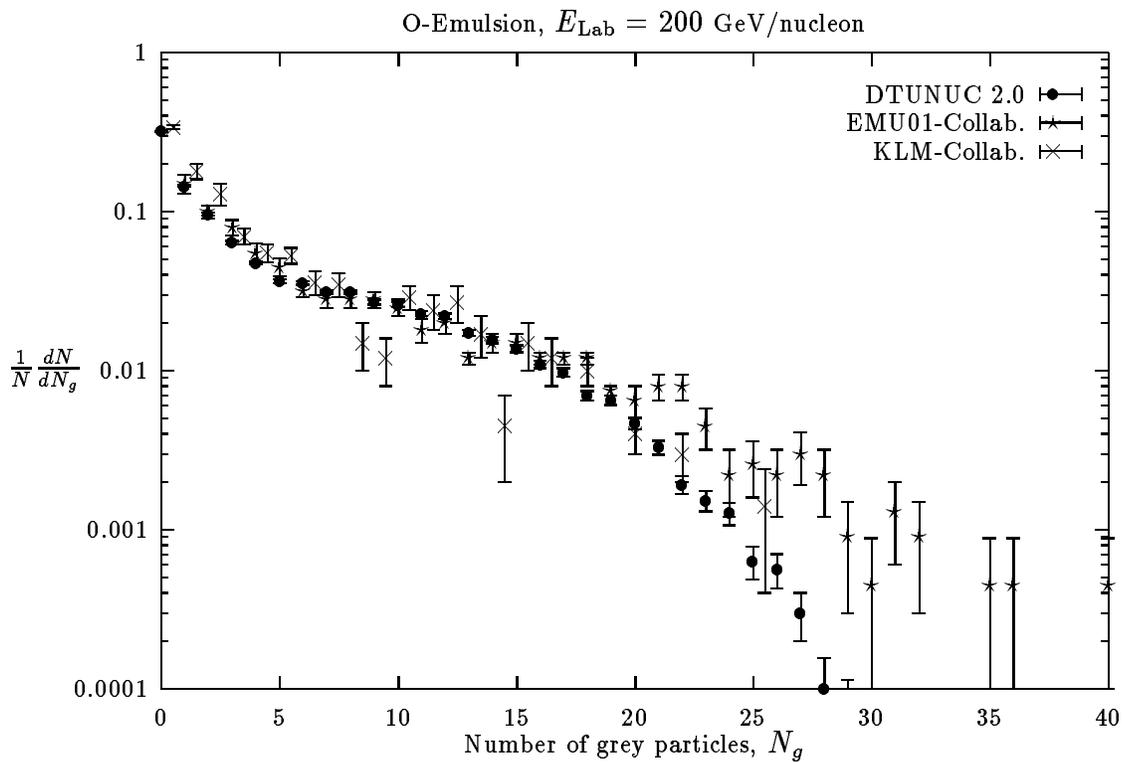
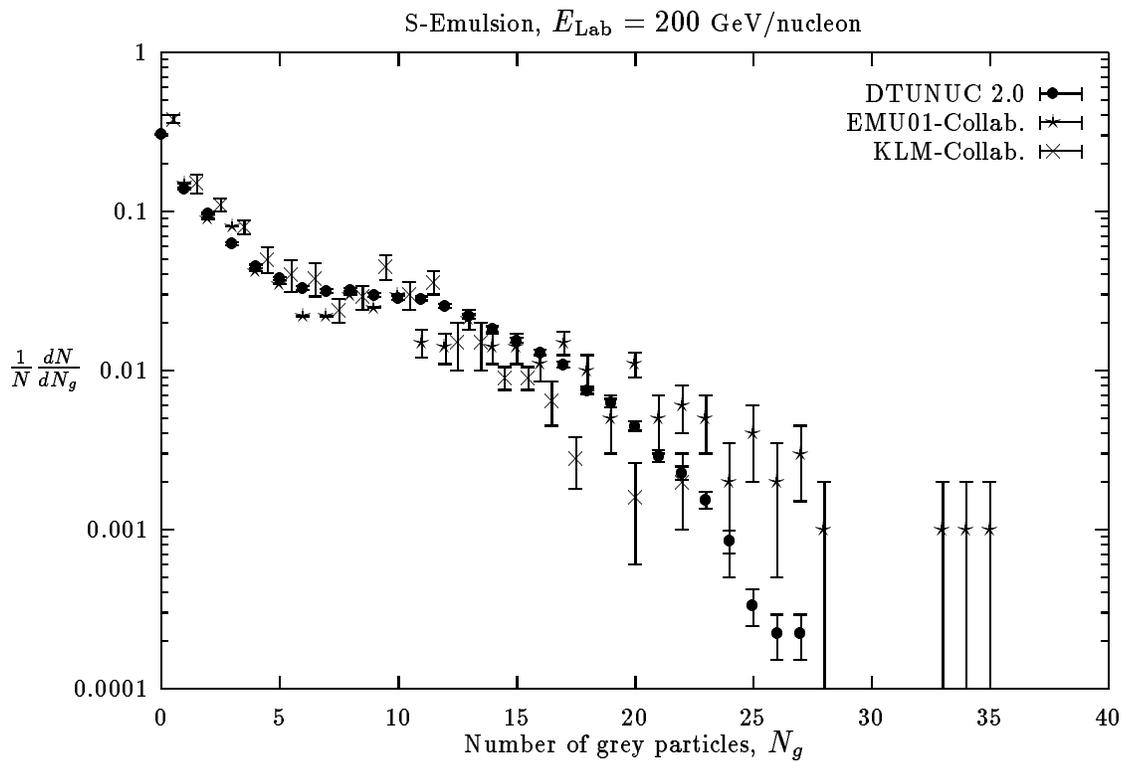
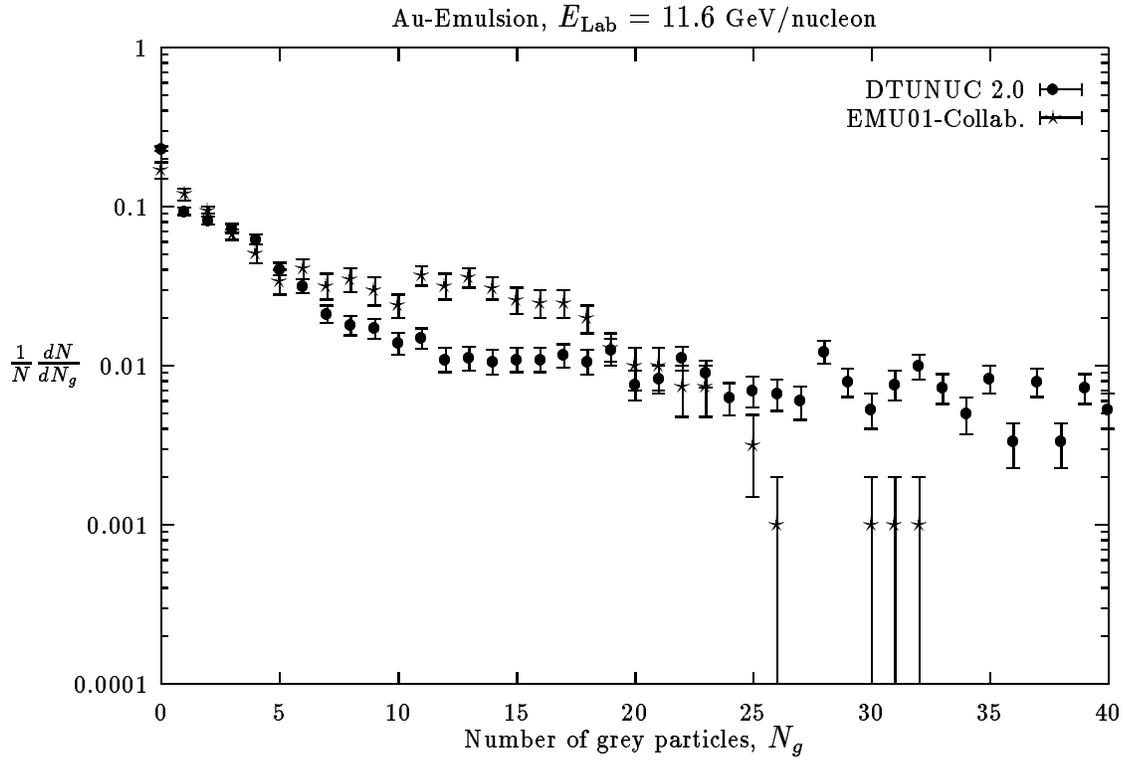
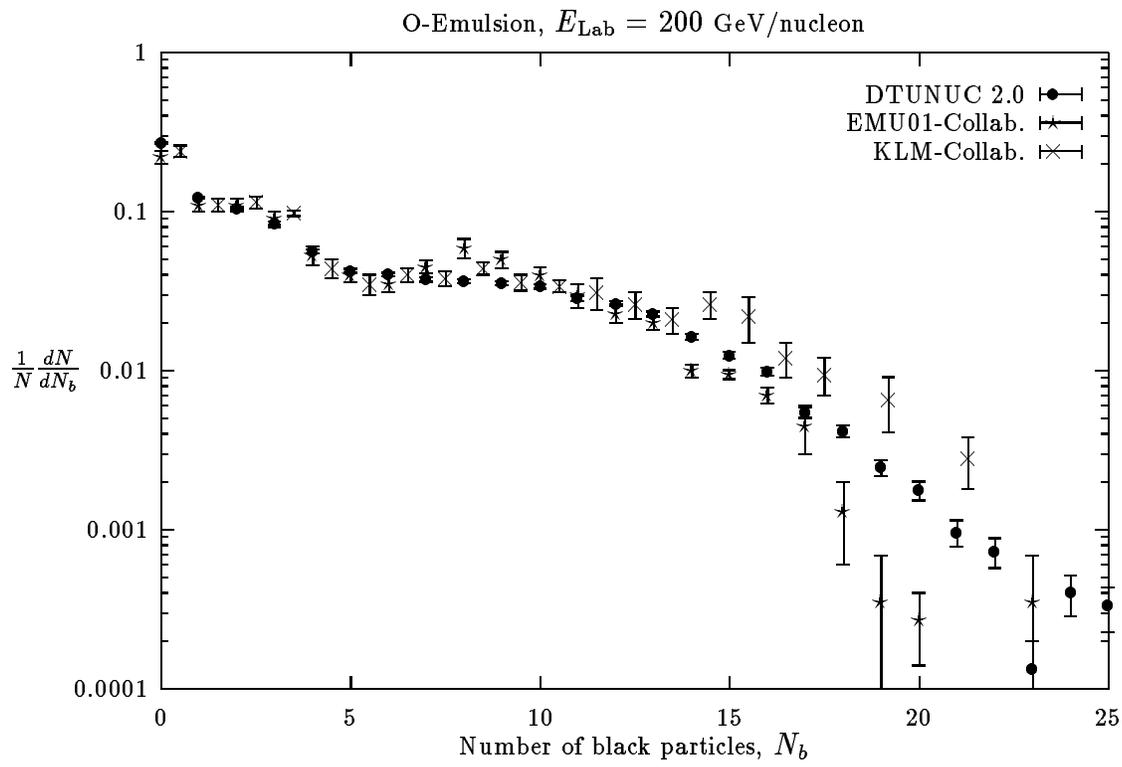
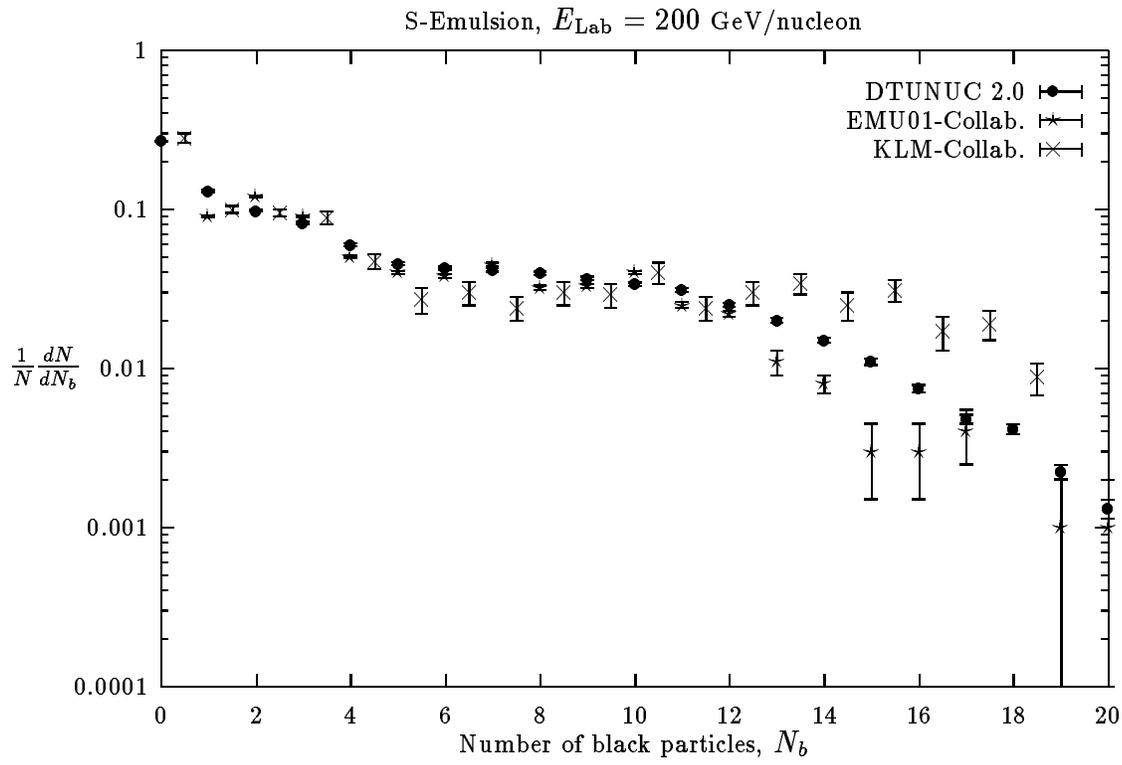
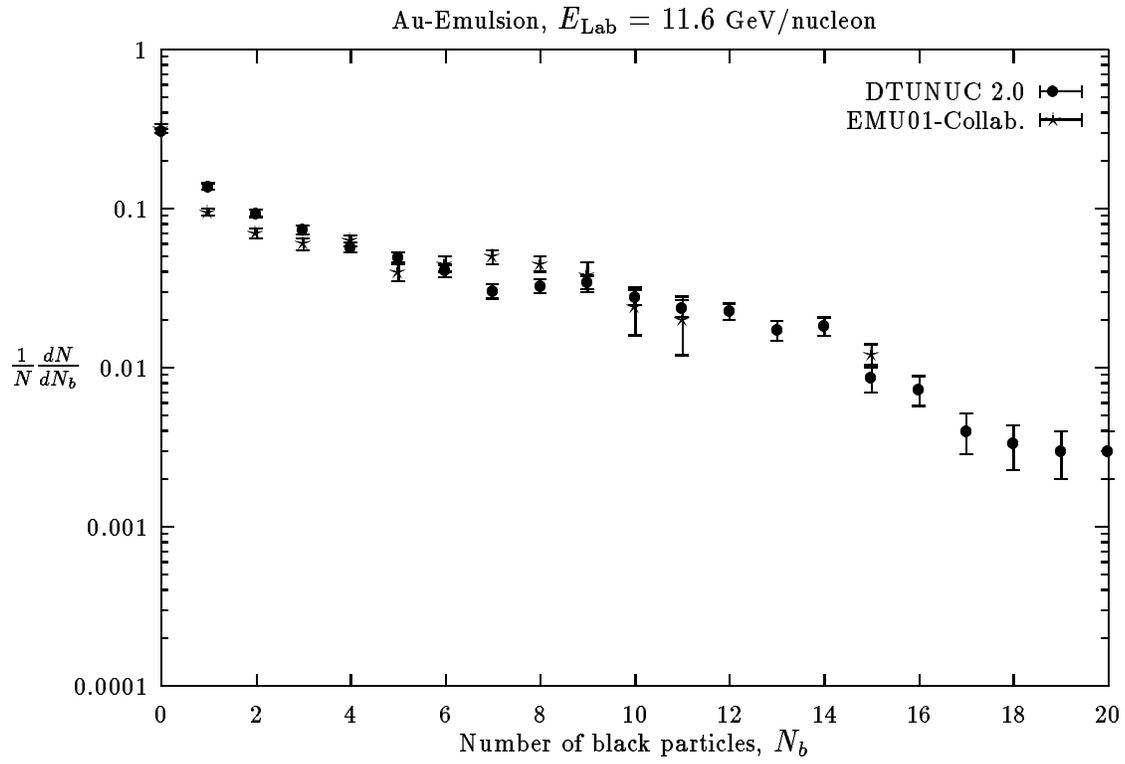
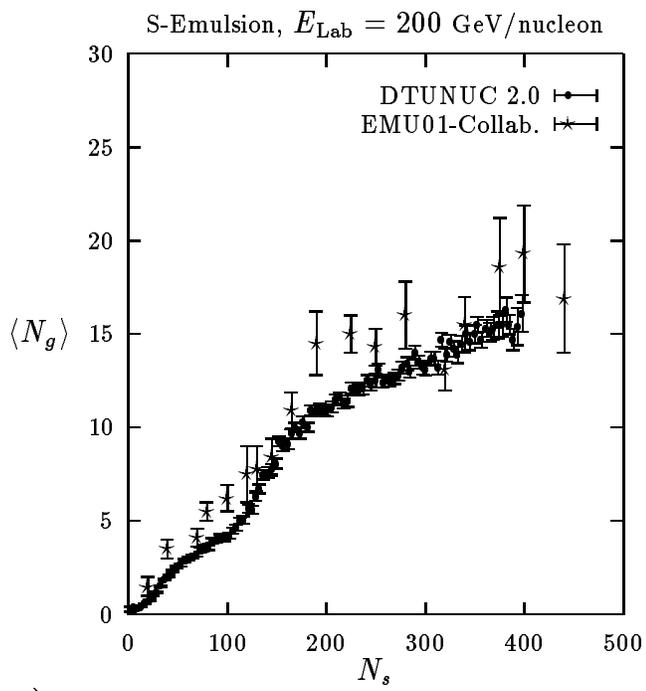
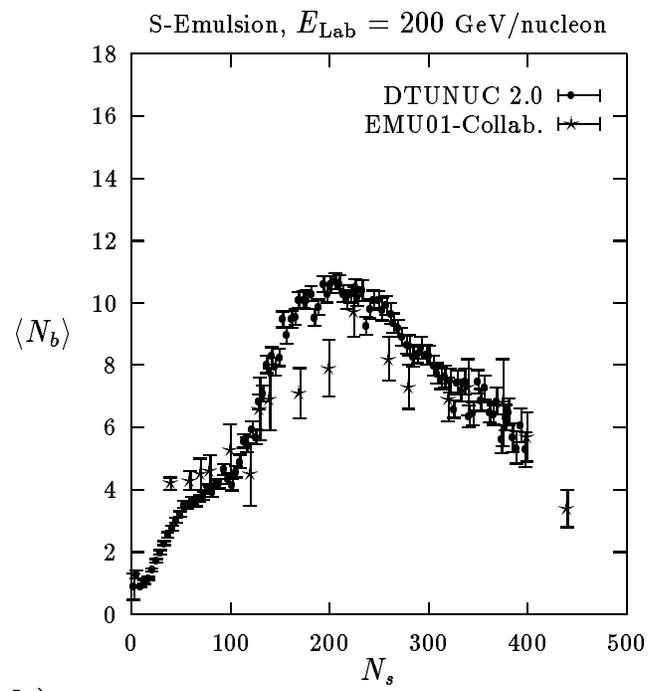
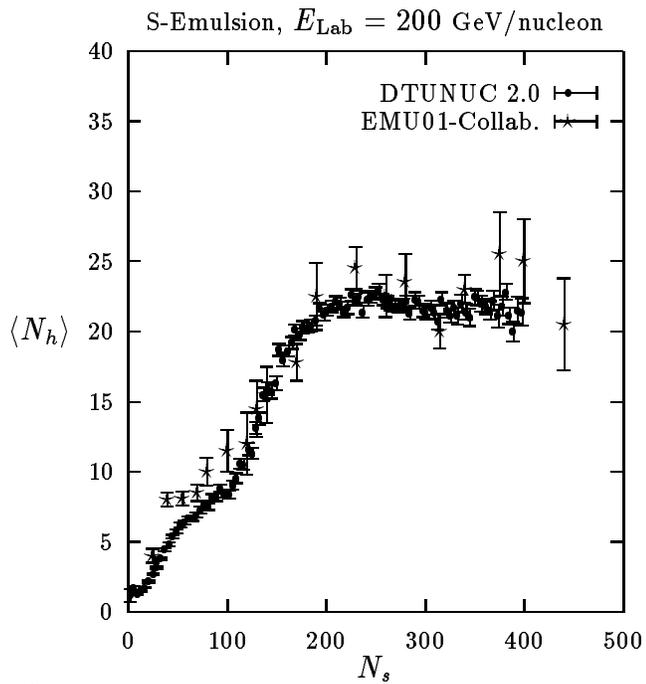
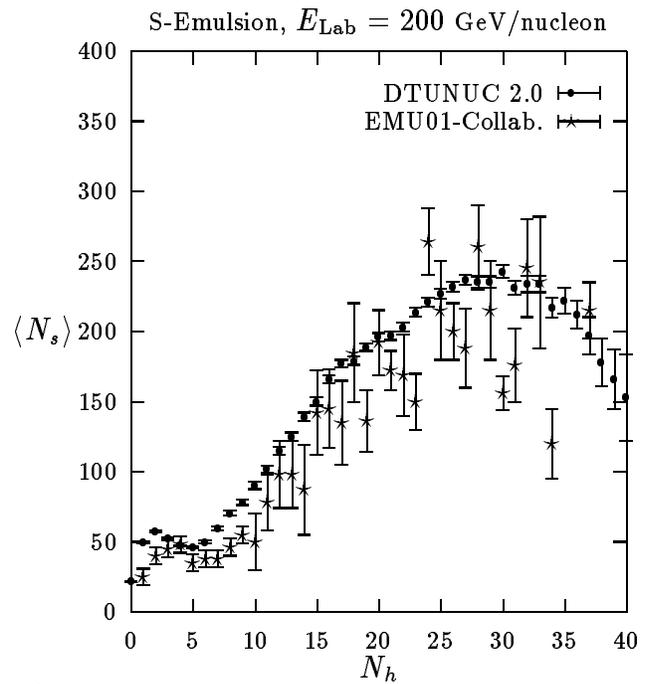
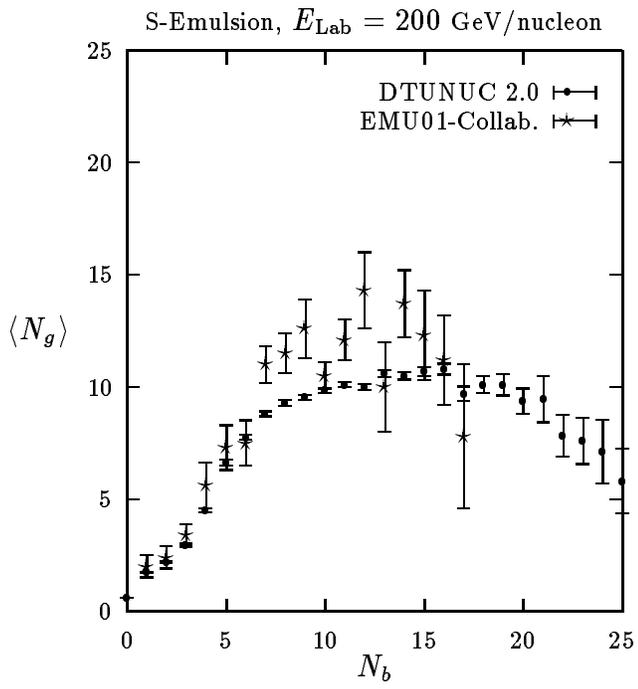
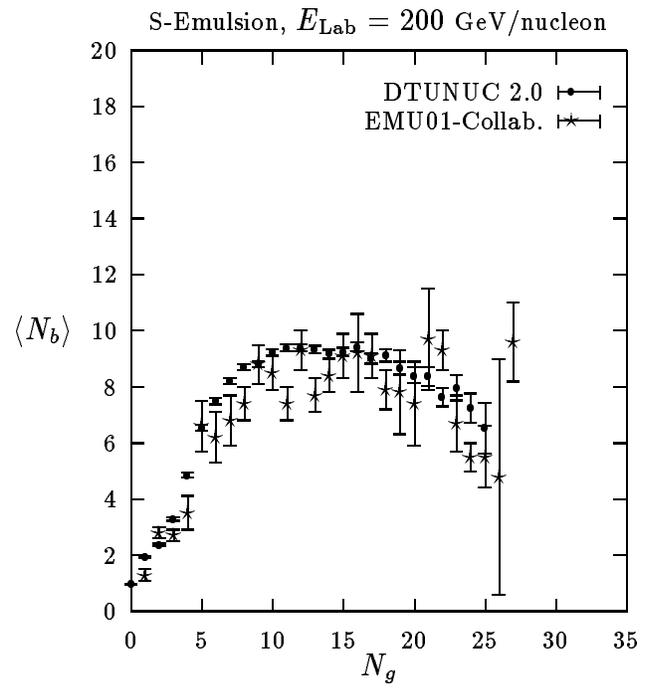
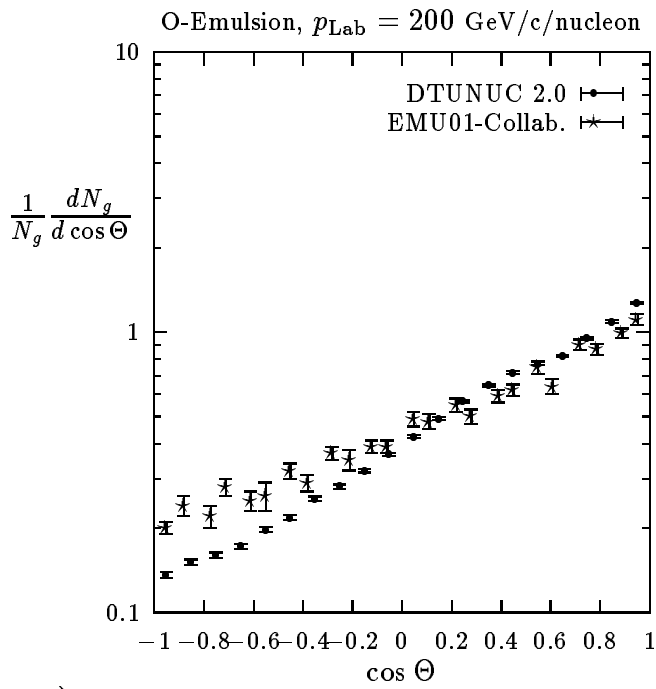
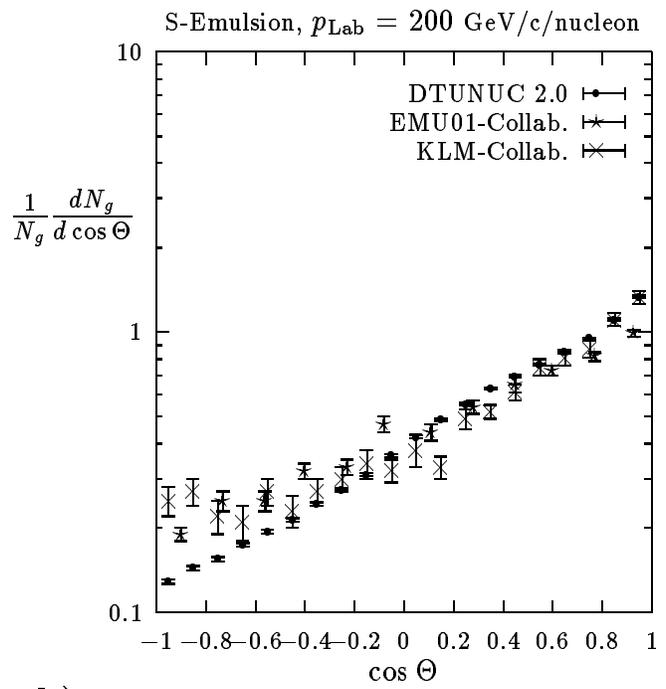
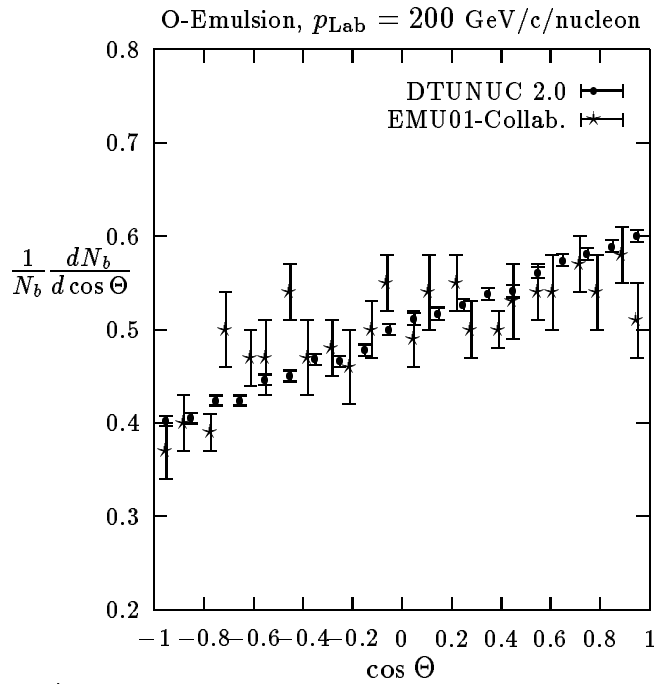
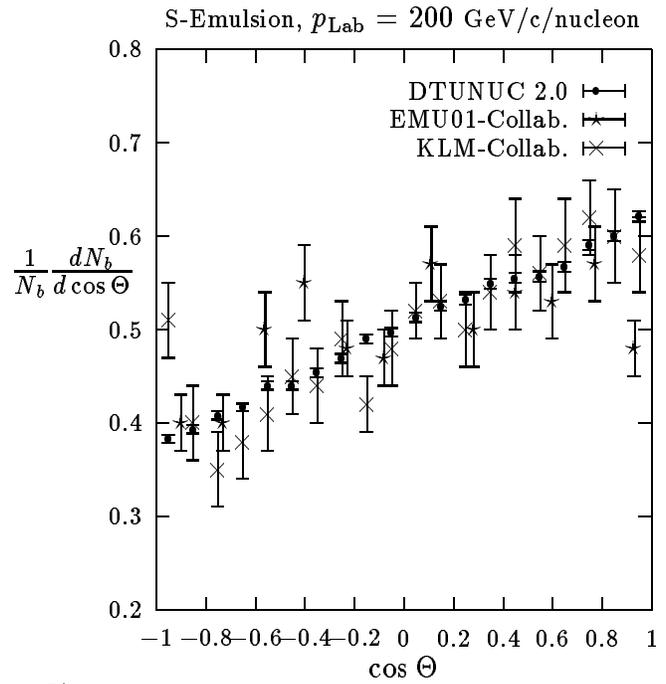
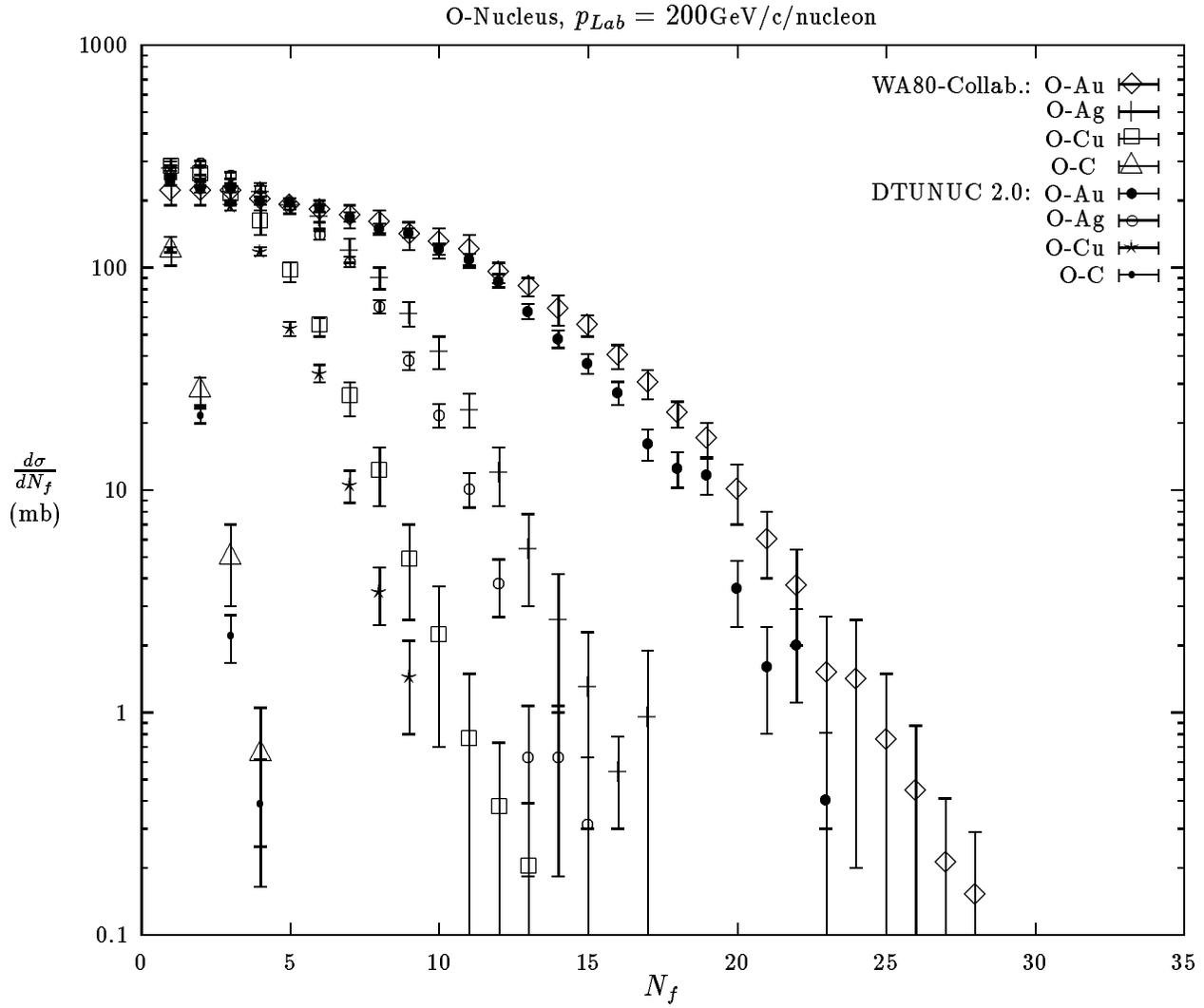
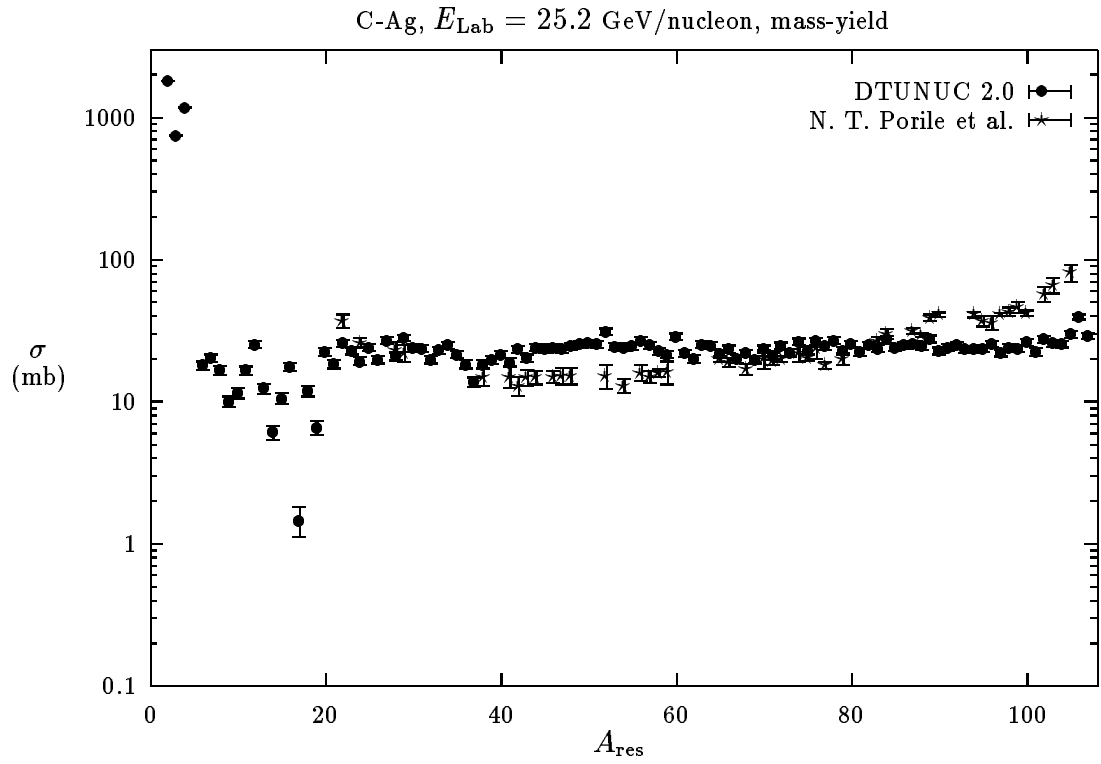
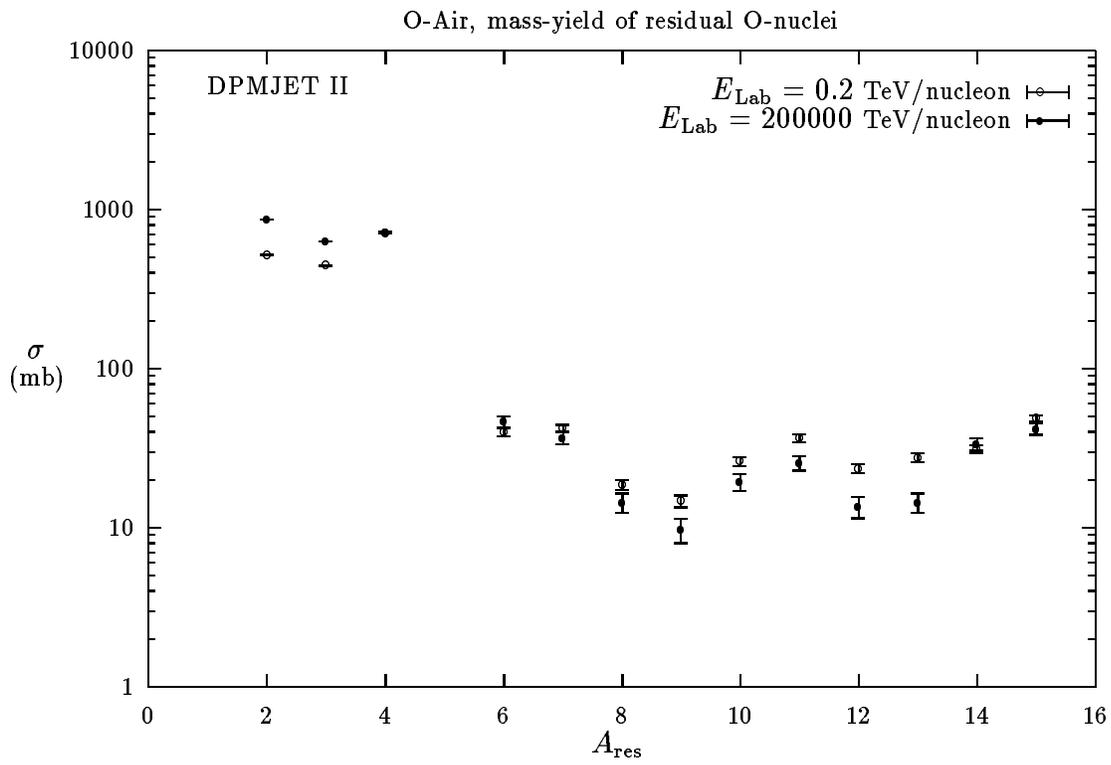
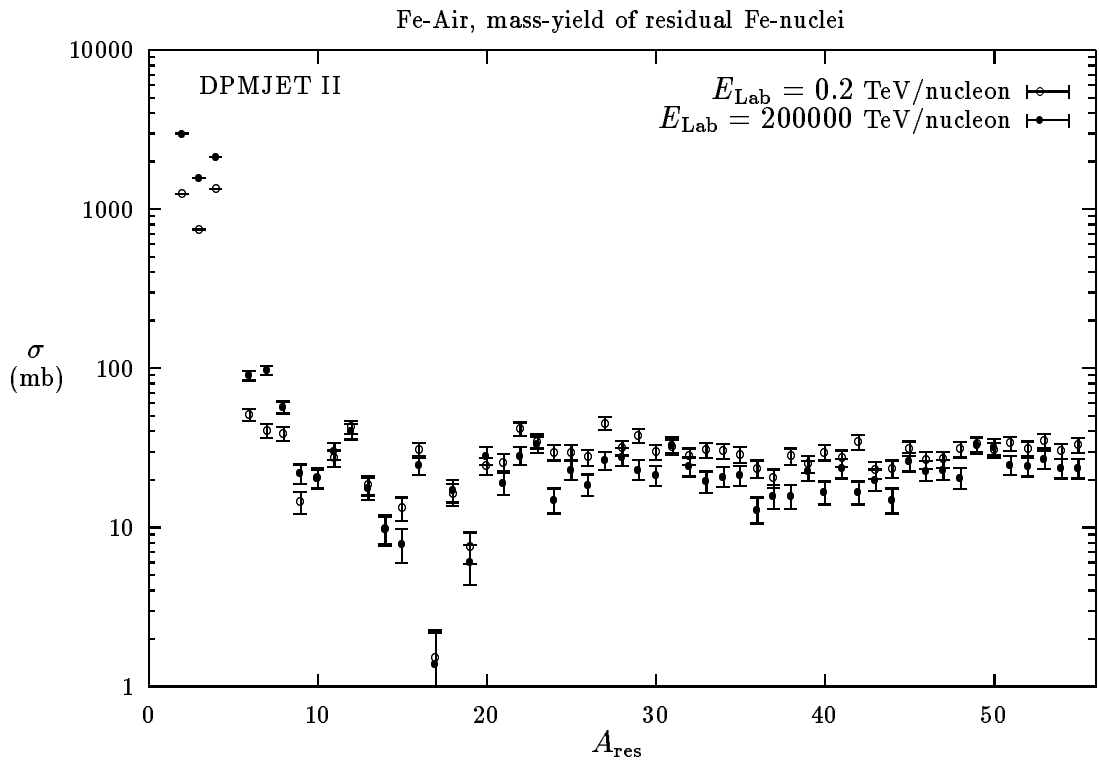
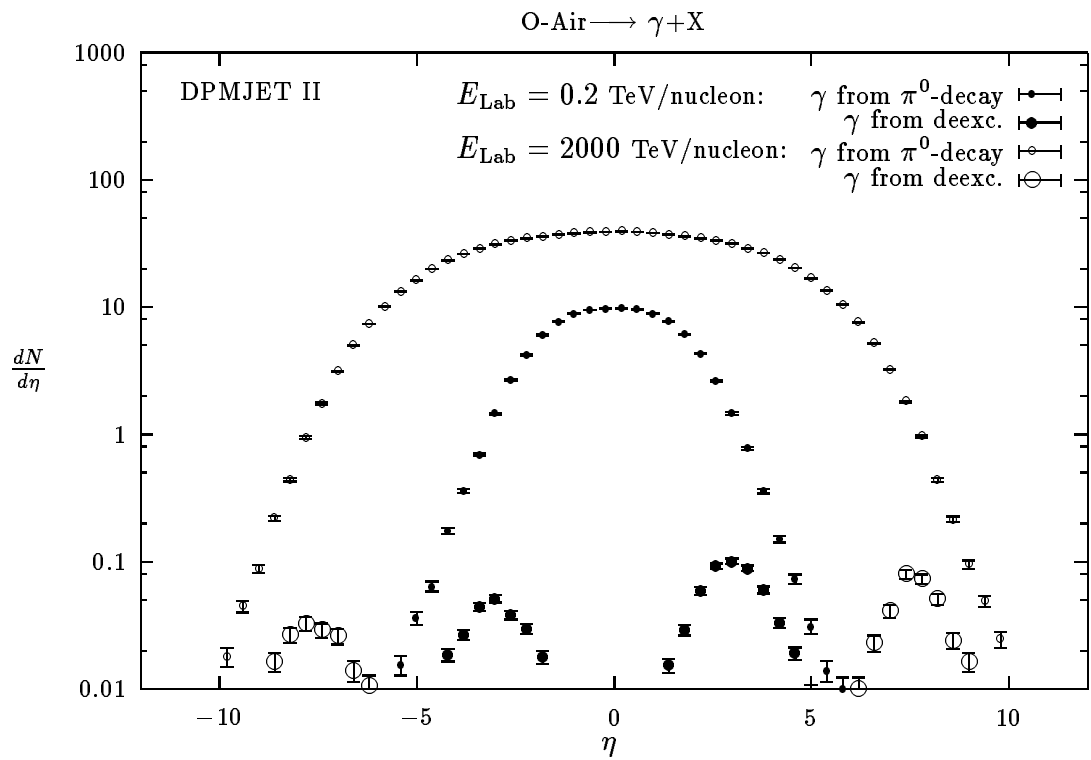
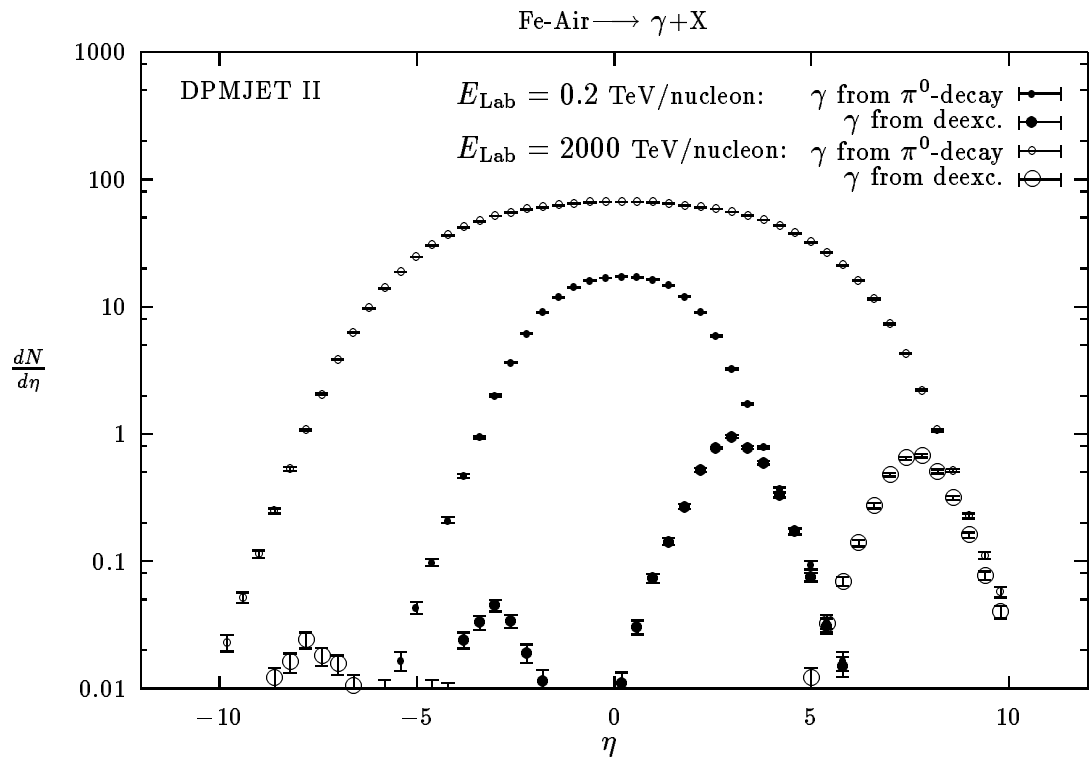
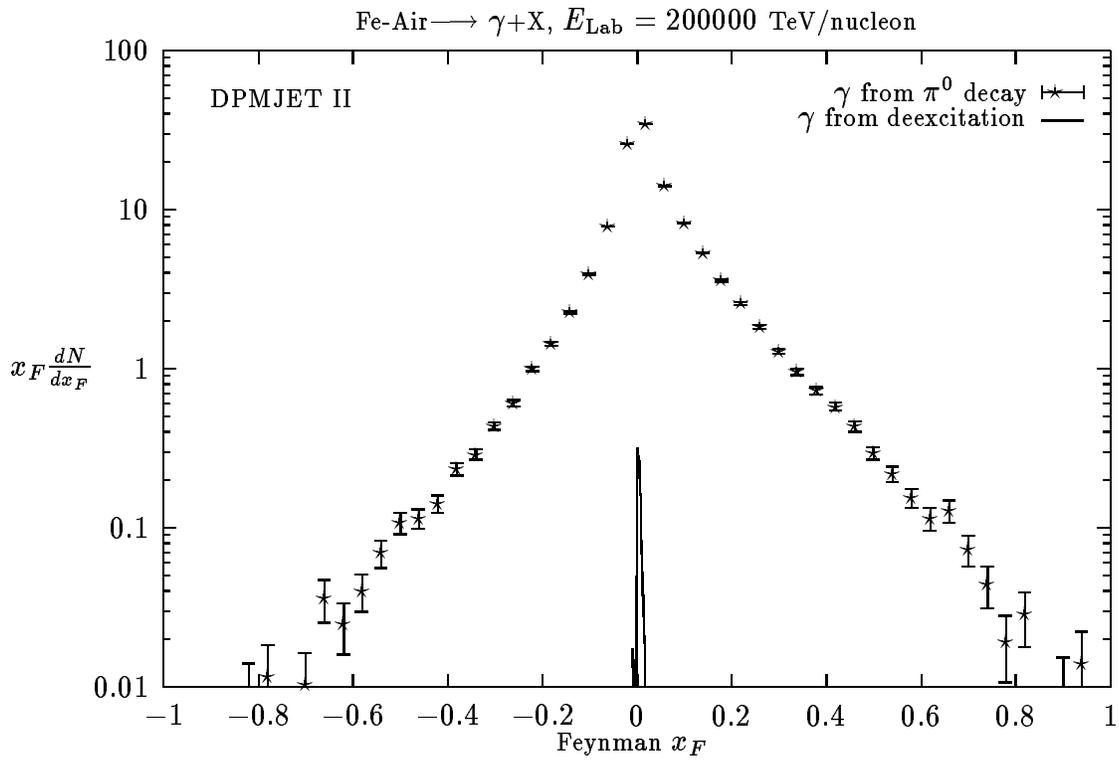
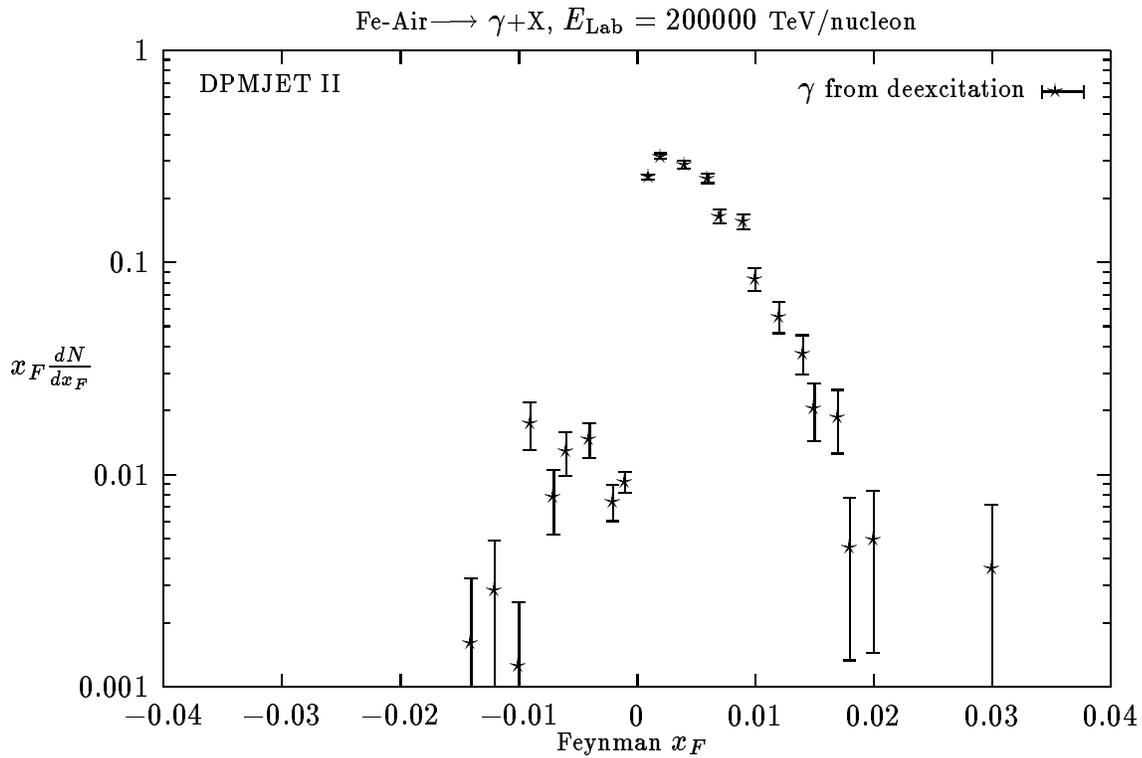
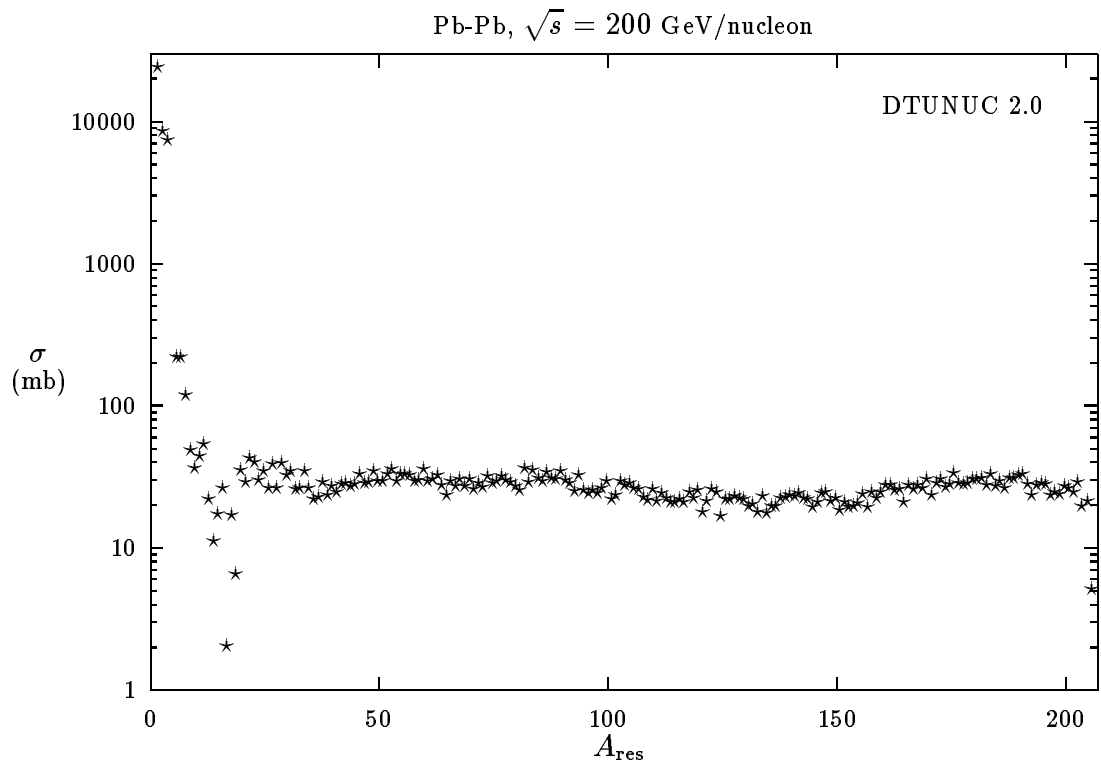
\end{document}